# Reexamination of Evaporation from Horizontal Surfaces with Implications for Solar Interfacial Evaporation Experiments

James H. Zhang[1], Rohith Mittapally[1], Guangxin Lv[1], Gang Chen[1*]

[1]Mechanical Engineering Department, Massachusetts Institute of Technology,

Cambridge, MA 02139

*Corresponding author: gchen2@mit.edu

**Abstract**

To explain reported solar interfacial-evaporation rates from porous materials beyond an apparent 100% efficiency using the thermal evaporation mechanism, many publications hypothesize that intermediate water inside porous materials have a reduced latent heat. Key supporting evidence is that water-only surfaces have lower dark evaporation rates than porous evaporators, with the ratio of the two rates taken as the latent heat reduction. Through simulations and experiments, we present benchmark evaporation rates of water and show that reported differences in natural evaporation are likely due to experimental error from recessed evaporating surfaces rather than from reduced latent heat**.** A few millimeters recession of the evaporating surface can drop evaporation rates over 50% due to a stagnant air layer, suggesting that the comparative experiments are prone to error and the latent heat reduction hypothesis cannot be substantiated. Our results indicate that new mechanistic directions need to be pursued to understand superthermal evaporation.

## INTRODUCTION

Driven by chemical potential difference and temperature gradients, evaporation is a ubiquitous phenomenon and has drawn intensive studies for applications such as micro-electronic cooling (*1*), triboelectric generators (*2*), critical mineral harvesting (*3, 4*), and solar desalination (*5–8*). Solar interfacial-evaporation technologies that use a porous black material on the water surface to absorb sunlight has drawn particular attention over the last decade (*5, 6*). Surprisingly, many groups have reported evaporation rates way beyond the thermal limit for single-stage distillation that is calculated based on the latent heat of water and incoming solar energy, i.e., superthermal evaporation (*9, 10*). Despite being superficially simple, the transport process is quite complex and there is ongoing debate about the mechanisms of superthermal evaporation under sunlight (*9–12*).

The originally proposed mechanism for superthermal evaporation rates, referred to in many subsequent publications, is that water has a reduced enthalpy inside of the porous interfacial evaporators (*9*). A key experiment to evaluate the reduced enthalpy is by comparing natural evaporation rates from a porous material to natural evaporation rates from a water-only interface, which is often referred to as "dark evaporation" in contrast with that under solar irradiation. It is often observed that the dark evaporation rates from porous surfaces are higher than that from water-only surfaces. By assuming equal heat input from the environment for both cases, the higher evaporation rate is attributed to the reduced latent heat of intermediate water inside the porous materials. Despite its wide acceptance, the validity of this approach had been questioned. Li et al. highlighted the many challenges that can be faced in measuring evaporation rates, such as container geometries and evaporating height recession (*13*). Previous works have also hypothesized that increased surface area due to microporosity inside of the material can potentially explain increased evaporation rates observed in dark conditions (*11, 14*). In this work, we will demonstrate that the apparent higher dark evaporation rates from porous materials are due to the reduced evaporation rates from water-only surfaces recessing below the lip of the container, rather than enhanced performance from porous materials due to the reduced latent heat of the intermediate water or increased surface area from microporosity.

In the limit of no external forced convection in the air, only natural convection occurs due to buoyancy effects arising from the temperature and vapor concentration gradients (*15, 16*). The convective flow pattern is of critical importance because it strongly influences the mass transfer resistance, impacting how quickly vapor molecules can escape the evaporating surface into the far field. In natural convection of water without additional heat input, both air and water typically circulate downwards because the surface fluid is colder and denser than the bulk fluid. These flow patterns in natural convection are very sensitive to the detailed geometry and significantly impact evaporation rates because these patterns govern both ambient heat input and vapor transport. The kinetics of vapor transport has been underappreciated in the field as a rate determining step in evaporation experiments.

The purpose of this work is to carefully analyze the dark evaporation rates expected from water-only surfaces and highlight the problems in using such experiments to interpret superthermal evaporation rates observed in solar interfacial evaporation. Finite element analysis (FEA) and experiments in controlled environments were conducted to benchmark the evaporation dependence on the container size, the ambient humidity, and most importantly, on the water level inside the container relative to the container edge. Our simulation and experiments show that water recessing below the top surface of the container by a few millimeters can reduce the evaporation rate from water-only surface by 50%. Furthermore, our work shows that increased surface area due to micron-scale porosity cannot lead to enhanced dark evaporation rates because the air side vapor concentration boundary layer is the dominating mass transfer resistance and that reduced latent heat does not lead to the higher dark evaporation rates reported due to vapor kinetics being the rate limiting step. Our analysis of literature suggests that most of the reported higher dark evaporation rates from porous materials are due to lower evaporation rates from recessed water-only surfaces rather than enhanced evaporation rates from porous materials. Thus, the dark evaporation rates comparison in the past literature is flawed. Extending the modeling to solar evaporation, we show that a reduced latent heat model cannot explain the high superthermal evaporation rates commonly reported in literature. A key picture missing in the reduced latent heat argument is that a cooling effect must occur when bulk water enters the porous material and forms an intermediate state. The sample-water interface absorbs additional heat to compensate for the enthalpy difference between bulk water and intermediate water. All the additional absorbed heat must be supplied from the environment. No solar superthermal evaporation could happen in this picture unless any part of the system stays below the environment temperature so that additional environmental heat can be absorbed. Such below-the-ambient cooling effect under solar radiation had not been observed from nearly flat 2D surfaces. These results invalidate the hypothesis of latent heat reduction of water in porous materials and call for exploration of other mechanisms such as the photomolecular effect (*10, 11, 17, 18*).

## RESULTS

**Comparison of FEA Simulations and Experiments on Natural Evaporation**. Simulations and experiments were conducted to benchmark dark evaporation rates from water-only interfaces exposed to the ambient (Fig. 1A). In the simulation, we modelled natural convection coupled with evaporation from water inside an open container in contact with a large air reservoir, representing the typical experimental setups in interfacial evaporation experiments. Navier-Stokes equation for flow in water and air, energy conservation equations of water, air, and the acrylic container, and mass transfer equation of water vapor in the air are solved simultaneously in transient simulations. The boundary conditions of open boundaries, constant ambient temperature, and constant vapor concentration on the outer boundary of the air domain were imposed on the model. Constant ambient temperature was also applied on the bottom impermeable wall of the domain on which the water container and air is in contact with. Simultaneously, we have also conducted experiments of water-only evaporation inside a large chamber with controlled humid air flow. During the experiment, we measured the evaporation rate from the water container as a function of ambient humidity and recessed surface height (see Methods section for complete details of the simulations and experiments).

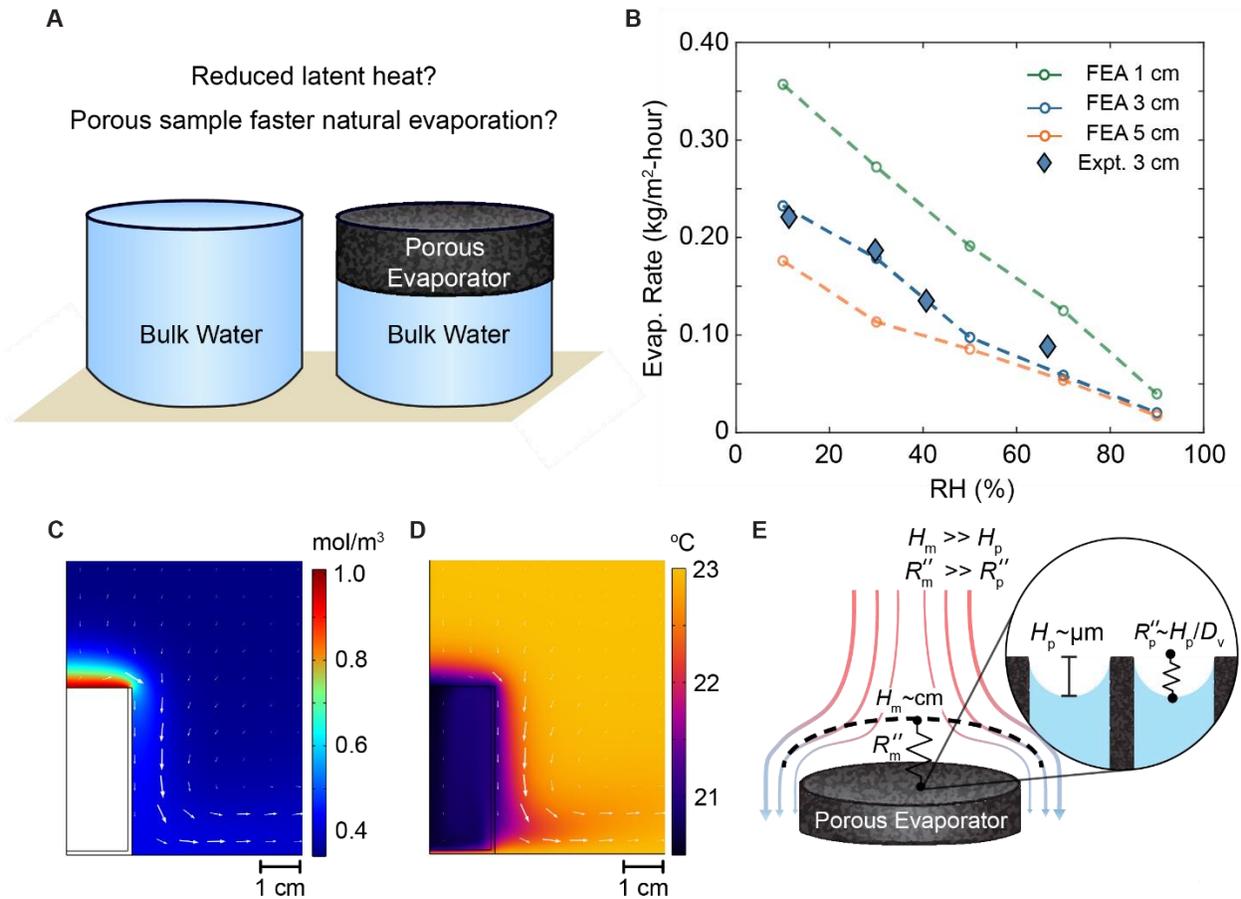

**Figure 1. Flow pattern and mass transfer during dark evaporation.** (**A**) Schematic of common experimental setup for comparative natural evaporation experiments between water-only interface and porous evaporators. (**B**) Predicted evaporation rates from the FEA model at different relative humidity (RH) values and container sizes. Comparisons with experiments are shown with solid diamonds. The ambient temperature is kept at 23 °C for simulations. For RH=30%, (**C**) vapor density distribution is plotted along with vapor flux as white arrows and (**D**) temperature profile along with natural convection-induced flow velocities as white arrows. Vector arrow lengths depict magnitudes. (**E**) Schematic of the mass transfer resistance inside wetted pores, $R_p''$, due to the meniscus relative to the mass transfer resistance from the vapor concentration boundary layer, $R_m''$.

The results for evaporation rates from the simulations and experiments on water are presented in Fig. 1B in which good agreement is found between the simulation and experiment for a container size of 3 cm diameter. As seen in Fig. 1B, the size of the sample strongly influences the evaporation rates. A container with a diameter of 1 cm will lead to evaporation rates 53% higher than that of a 3 cm container at 10% relative humidity. The evaporation rate decreases almost linearly with the ambient relative humidity due to the smaller vapor mole fraction difference between the surface and ambient.

To further understand where the heat comes from, the average heat flux at the evaporating surface is analyzed in the simulations for a 3 cm diameter container exposed to 30% RH (Fig. S1). Most of the heat for evaporation is provided from the ambient environment through the container walls and into the bulk liquid below the evaporating surface, in agreement with a previous study (*14*). This is due to the larger surface area of the container sidewalls and the convection flows inside of the water. The density stratification of water is unstable due to it being colder on top, making it very effective at providing heat from the ambient air around the sidewalls of the container to the top water-air surface for evaporation. It is estimated that a 0.0012 °C difference between the top evaporating surface and the bottom of the container is sufficient to cause convective mixing in the water (see SI Note 1). The heat flux provided from radiation and air side are lower and have similar values. Since most of the heat is provided through the liquid pathway, it can be expected that the inclusion of a porous material would affect heat transfer from the bulk liquid due to their typical thermally insulating features, leading to changes in the evaporating surface temperature and overall heat balance.

**Flow Pattern and Concentration Boundary Layer.** The patterns for both mass transfer and heat transfer on the air side were analyzed to understand the boundary layer. The vapor concentration boundary layer (Fig. 1C) and the thermal boundary layer (Fig. 1D) above the evaporating surface has similar characteristics. The thickness of both boundary layers is about 1 cm in the center region due to stagnation and their height tapers off near the edges of the surface. Fig. 1C also illustrates the magnitude and direction of vapor flux (represented by white arrows). The vapor flux is due to combined effects of both diffusion and natural convection. The flux of water vapor is fastest at the edge of the evaporating surface as shown by the larger white arrows, flowing around the edge and downwards. The reason why it flows downwards is revealed in Fig. 1D, which shows the temperature distributions as well as the convection flow patterns, represented by the white arrows. The sharpest temperature gradient in the air side is also near the edge of the container and the air is colder around the container. The dense cold air will sink, causing the natural convection air current to flow downwards. The coupling between the concentration distribution and temperature distribution leads to the vapor to convectively flow downwards and water to evaporate faster near the edges. This effect explains why smaller samples evaporate faster because the perimeter to area ratio becomes more favorable, leading to stronger edge effects.

The mass transfer resistance of vapor into the air across the boundary layer can be estimated from the simulations using the general mass transfer convective boundary condition

$$\dot{m}''_{evap} = \frac{M_v C_g \left(c_{v,s}(T_s) - (RH)c_{v,s}(T_\infty)\right)}{R''_m} \quad (1)$$

where $\dot{m}''_{evap}$ is the mass flux per unit area, $R''_m$ describes the convective mass transfer resistance of water vapor through the air per unit area between the evaporating surface and the ambient, $T_s$ is the surface temperature, $T_\infty$ is the ambient temperature, $c_{v,s}$ is the saturated vapor mole fraction at a given temperature, $RH$ is the ambient relative humidity, $C_g$ is the molar density of ambient air, and $M_v$ is the molar mass of water. Using the simulation data, it was found that $R''_m$ ranges from 114 s/m to 373 s/m for the studied different container sizes and ambient humidity values.

The thickness of the boundary layer provides a length scale that characterizes the evaporation mass transfer resistance on the air side. For example, many reported interfacial evaporators have surface areas larger than its projected surface area through either open pore structures or 3D macrostructures (*19–21*). If the surface structure is much smaller than the boundary layer, i.e. with micron-sized pores, then the additional surface area may not contribute to evaporation as depicted in Fig. 1E. Water vapor needs to diffuse from the enhanced surface area in the meniscus out to the far field. Although there is a larger evaporating surface area, the water vapor does not have the kinetics to escape from the micron-pores into the far-field outside of the boundary layer. This is because this diffusion resistance inside the pore meniscus is in series with the convective boundary layer resistance outside, resulting in the rate limiting step to be through the macroscopic concentration boundary layer. The mass transfer resistance inside the pore can be estimated using a simple diffusion expression

$$R_p'' \approx \frac{H_p}{D_v} \qquad (2)$$

Using a micron height pore and the diffusion coefficient of vapor in air at ambient temperatures of about 0.24 cm$^2$/s, $R_p''$ is about 0.04 s/m, which is four orders of magnitude smaller than the boundary layer resistance as found from Eq. (1). As a result, the region near the pore remains close to the saturated vapor pressure condition and it cannot explain enhanced evaporation from microporous materials. If the characteristic size of the surface roughness or macrostructures is comparable or larger than the boundary layer thickness, then the additional surface area may contribute to evaporation, as seen in open porous materials with higher evaporation rates due to forced convection or 3D evaporating macrostructures (*19–21*).

Two effects not discussed here that could enhance evaporation kinetics is Stefan-flow induced vapor transport and increased surface vapor molar concentration from pressurized water in hydrophobic nanopores. Stefan-flow can increase the vapor transport kinetics in the air side due to an induced Stefan velocity, leading to a "blowing-effect" at the evaporator interface (*16*). However, this effect is only noticeable at elevated surface temperatures (> 60 °C) when the vapor mole fraction becomes comparable with that of air (*22*). Pressurized water inside of hydrophobic nanopores will have an elevated surface vapor concentration due to the Kelvin effect from positive meniscus curvature, leading to enhanced evaporation rates. This effect is unlikely in natural evaporation experiments because it requires the material to be very hydrophobic, have nanopores, and the water to be pressurized many atmospheres above the ambient inside the testing container (*23*).

**Water Recessed Height**. The above discussion of the flow pattern allows us to discern four potential pitfalls in past dark experiments used to compare evaporation rates from water-only and from porous surfaces (Fig. 2A-C). First, for water-only experiments, the water surface will be recessed inside the container if a beaker is used to avoid water overflowing into the pouring lip, as shown in Fig. 2A. Porous evaporators do not have this issue as they can be situated higher up inside the beaker or fixed in position due to expansion when it uptakes water, leading to experimental artifacts in relative height differences of evaporating surfaces. Secondly, since some porous evaporators are free floating on top of a water surface during experiments, the free-floating evaporator will be higher up if the same amount of water is used between tests. If the container doesn't closely match in size with the evaporator, the sidewall can also contribute to additional evaporation as illustrated in Fig. 2B and highlighted in the past (*13*). Thirdly, due to the irregular shape of many porous evaporators, it has become common practice to use a cylindrical container that is larger than the sample and cover the excess water surface area with foam. The sample will be in level with the foam during evaporation tests. However, the comparative water test will have a recessed level due to the thickness of the foam and artificially deflate the water evaporation rate as shown in Fig. 2C. Fourth, during longtime experiments, the water level from a free surface will drop from evaporation unless a connected reservoir is used. As a result, the apparent evaporation rate would be lower than an experiment where the height is maintained.

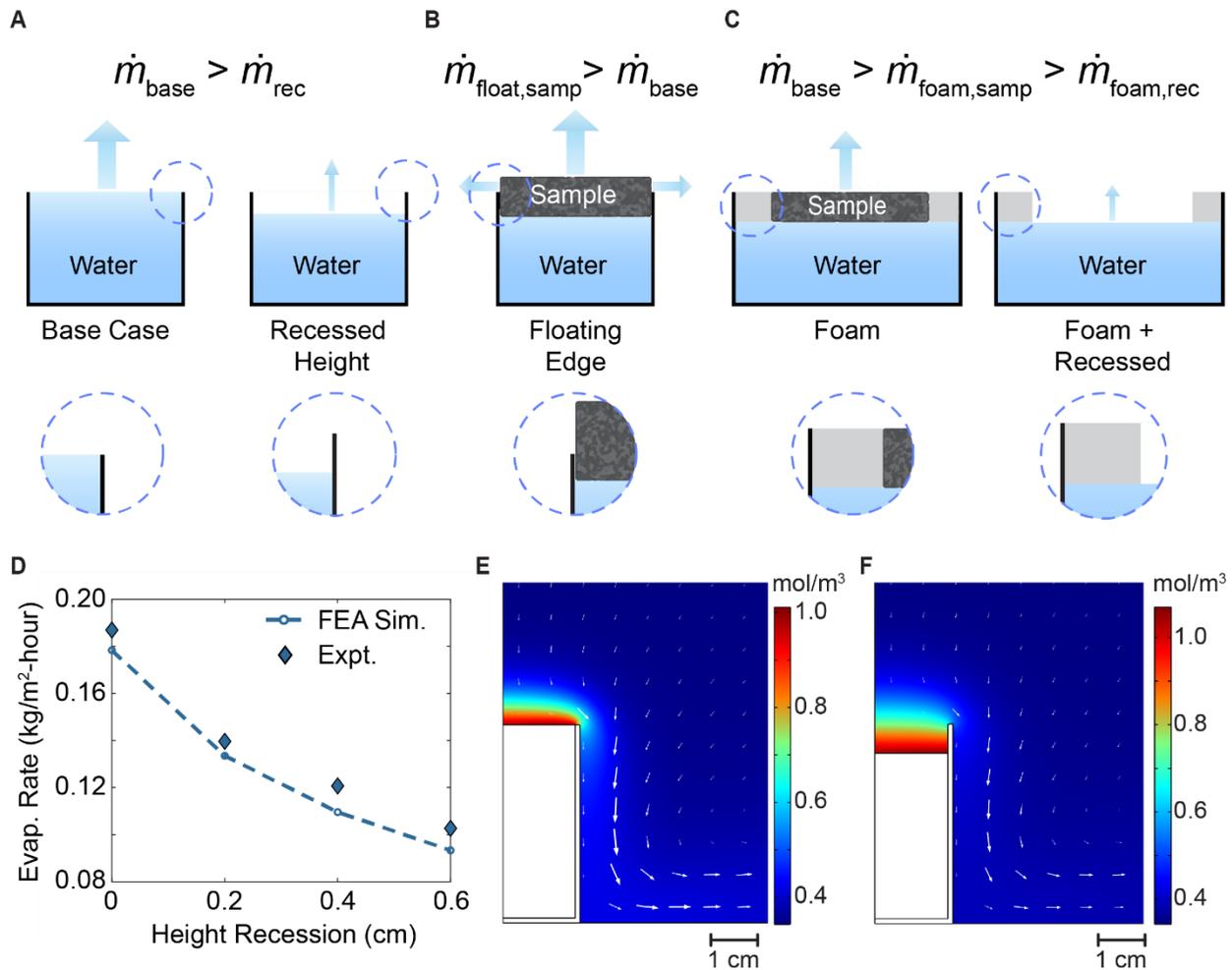

**Figure 2. Effects of experimental setup on apparent measured evaporation rates.** A) $\dot{m}_{base}$ is the base case rate when pure water is levelled with container lip, $\dot{m}_{rec}$ is for water recessed below the lip. B) $\dot{m}_{float,samp}$ represents the case when porous floating sample top surface rises above the container lip. Evaporation from the side leads to higher apparent rate than $\dot{m}_{base}$. C) $\dot{m}_{foam,samp}$ is the rate when foam insulation wraps around the sample during evaporation experiments, which is lower than $\dot{m}_{base}$ as the edge effect is reduced in the former. $\dot{m}_{foam,rec}$ represents the rate with foam in the absence of sample, which is even lower due to the recessed water level. D) Effects of recessed evaporating surface on evaporation rates from FEA simulations (blue dashed line) and experiments (blue diamonds). Simulations are conducted for 30% RH at 23 °C. Experiments are conducted in 31.1 ± 1.1% RH at 24.58 ± 0.15 K. FEA snapshots of the vapor concentration distribution and natural convection velocity magnitudes evaporating into 30% RH and 23 °C ambient with the water surface E) levelled with the container lip and F) 6 mm below the container lip. Arrow lengths are normalized linearly to the largest velocity seen in both simulations.

Among the above discussed scenarios, recessed water level from the top exists in pretty much all previous experiments, and it is the most problematic because a dropped height will create a stagnant air film above the evaporating surface due to the stable density stratification, thereby decreasing the air flow velocities and the evaporation rate. Through a combination of simulations and experiments, we intentionally recessed the evaporating water surface from the container lip to quantify its effects on evaporation rates as summarized in Fig. 2D. A 2 mm recession will cause the evaporation rate to drop by 25% and a 6 mm recession will cause it to drop by 47.7% This stagnation layer can be seen clearly by comparing FEA snapshots of the natural convection flow patterns and vapor concentration gradients in Figs. 2E and 2F, the former with water surface aligned with the lip of the container, while the latter with water surface recessed. As a result, water vapor molecules must first diffuse through the stagnant layer between the water surface and the lip of the container, which leads to a mass transfer resistance of about 167 s/m as estimated using Eq. (2) with the 6 mm recessed height. This value is comparable and in series with the convective mass transfer resistance seen from earlier, causing the evaporation rate to drop by almost 50%. The coupling nature also makes the natural convective currents much weaker outside of the surface. In the comparisons between the two cases shown in Figs. 2E and 2F, it was found that the maximum natural convection velocity in air drops by 35% from 21.2 mm/s to 13.7 mm/s. From these results, even small height differences can lead to large changes in evaporation rates.

**Correlation Model to Study Porous Material Effects.** To reduce the dependency on FEA simulations and provide a simplified model for researchers, we have also developed a model using heat and mass transfer correlations and used it to understand the effects of porous materials on evaporation from horizontal surfaces. Our correlation model setup has similar characteristics as Caratenuto et al. (*14*), but we include the physics of mass transfer and more proper treatment of heat transfer inside of the water to give the model predictive power of evaporation rates. The full details of the model for water-only evaporation are described in SI Note 2 and the corresponding heat and mass transfer resistance diagrams are illustrated in Fig. S2.

We first validated the correlation model by evaluating the evaporation of water from the water-only interface and we found very good agreement in predicted evaporation rates with the simulation data as well as the experimental data (Fig. S3**)**. Using the correlation model, we can estimate the sensitivity of evaporation to different parameters to understand the effects of experimental setup (Fig. S4). We set the base case as water-only inside a container with a diameter of 3 cm and height of 4 cm that is evaporating into the ambient at 30% RH and 23 $^o$C. By testing different inputs into the model, we found that the evaporation rate is most sensitive to the ambient temperature with an evaporation rate changing by about 1.43% with every percent change in temperature in $^o$C. This is because the ambient temperature impacts the ambient vapor content at constant RH and changes the amount of heat it can provide to the sample during evaporation tests. The model also shows that natural evaporation rates have reasonable sensitivity to the ambient relative humidity at -0.52% and diameter at -0.40%. The evaporation rate is relatively insensitive to the height (-0.002%) and thermal conductivity of the container (-0.0005%). Although the heat transfer through the sidewalls of the container could potentially increase as the container walls gets thinner and taller, the mass transfer resistance of the water vapor in air cannot change significantly in natural evaporation conditions. As a result, the kinetics of vapor is the rate-limiting step.

We also used the developed correlation model to test the following two hypotheses related to evaporation from porous interfacial evaporators. The first hypothesis is that a reduced latent heat causes the porous evaporator to have higher evaporation rates under natural convection, as hypothesized in many papers in the field (*9, 24–29*). The second hypothesis is that a colder evaporating surface from porous materials causes faster evaporation rates (*14*). Using the developed model to predict evaporation from water surfaces, we extended the correlation model to describe evaporation from porous samples in natural convection conditions (see SI Note 3 and Fig. S5).

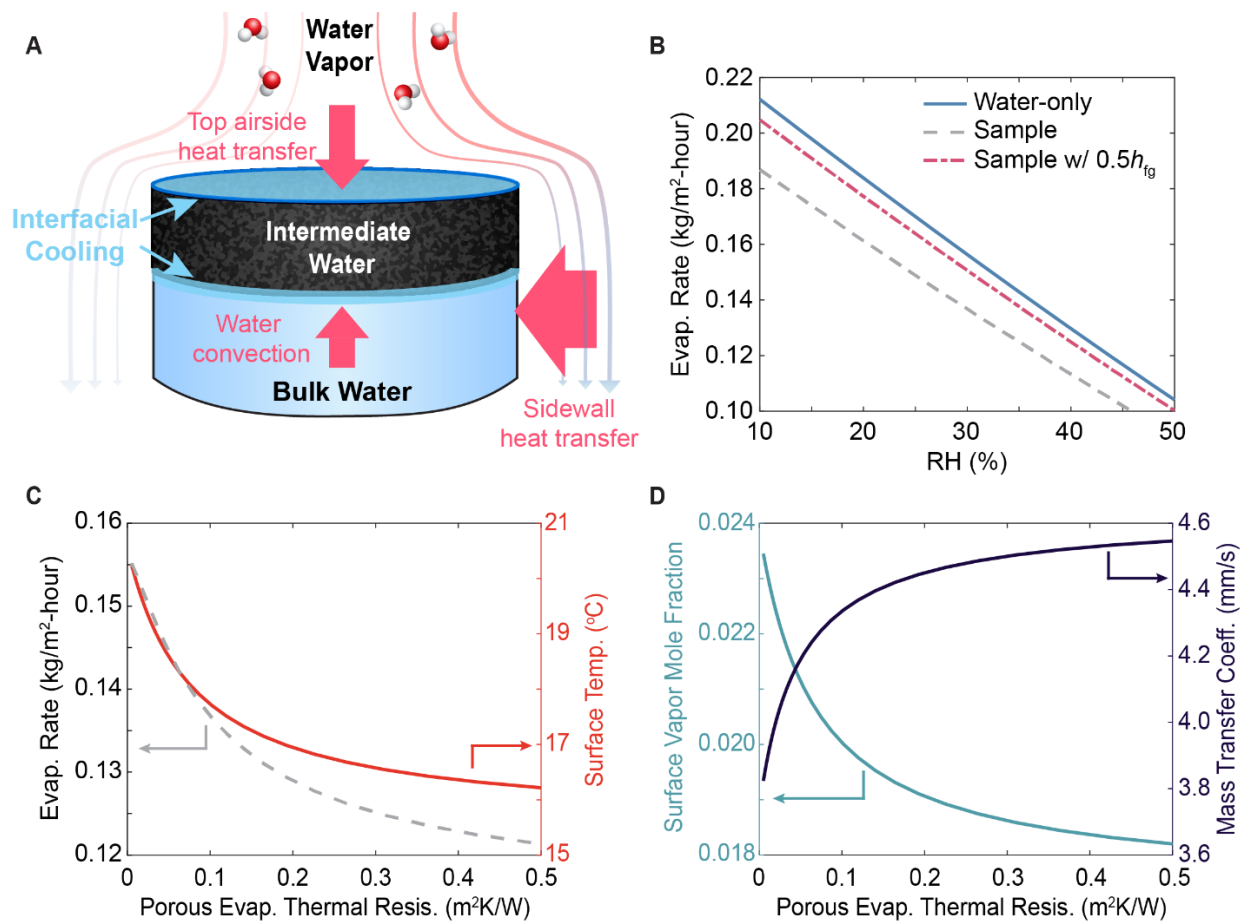

**Figure 3. Dark evaporation rates and surface temperature.** (**A**) Diagram of hypothesized water states inside of porous materials during natural evaporation experiments. An additional cooling effect must occur as bulk water enters the porous material from below if the intermediate water has lower latent heat. Differences in water enthalpies determine cooling at each of the interfaces. (**B**) Predicted evaporation rates as a function of relative humidity for water-only interface (solid blue line), sample with no water latent heat changes inside (dashed gray line), and sample with water inside having half of the latent heat of bulk water (dashed dotted red line). The thermal resistance of the porous evaporator is set at 0.1 m²K/W, ambient temperature of 23 °C, and sample diameter of 3 cm. (**C**) Predicted evaporation rate, surface temperature, (**D**) water vapor mole fraction on surface of evaporator, and mass transfer coefficient in air from sample evaporation with no reduced latent heat effect as a function of sample thermal resistance. The ambient RH is set to 30%.

**Reduced Latent Heat Doesn't Lead to Higher Natural Evaporation Rates than Water.** The introduction of a porous material with reduced latent heat will have two effects on the overall heat transfer in natural evaporation (Fig. 3A). The first effect is interfacial cooling on both the top and bottom-side of the porous evaporator. Due to energy conservation, the difference in enthalpies between bulk water and intermediate water inside of the porous sample must create a cooling effect at the bottom interface to satisfy energy conservation. Similarly at the top interface, the difference in enthalpies between the intermediate water and water vapor creates an evaporative cooling effect. Unless large amount of evaporated water is in cluster form, the sum of the two cooling powers at the two interfaces equals to that of the latent heat difference of pure water. The second effect that the porous material will create is a thermal insulation effect on heat transferred through the bottom pathway. The porous evaporator's thermal resistance per unit area, $R''_{samp}$, is defined as

$$R''_{samp} = \frac{t_{samp}}{k_{samp}} \quad (3)$$

where $t_{samp}$ is the thickness of the sample and $k_{samp}$ is the thermal conductivity of the wetted sample. Many interfacial evaporators are made from polymeric materials, which typically have thermal conductivities on the order of 0.1 W/m-K and water has a thermal conductivity of 0.6 W/m-K. The thermal conductivity of the interfacial evaporator and water mixture would have an intermediate value between the two extremes depending on the water and evaporator's mass ratio during experiments. Sample thicknesses range from mm to cm range, leading to $R''_{samp}$ to be on the order of 0.001 to 0.1 m²K/W range.

We consider three cases when calculating the evaporation rates (Fig. 3B). The first case is the water-only evaporation rate, as illustrated by the blue line. The second case is the sample evaporation rate shown by the dashed gray line, representing evaporation without reduced latent heat effects. The third case is the sample evaporation with the latent heat reduced by half inside of the porous evaporator, as illustrated by the dashed-dotted red line. All three curves have similar features. The evaporation rate is high at low ambient relative humidity values and drops almost linearly as the humidity increases. More importantly, the model predicts that the evaporation rate of the water-only interface should be the highest. At 30% RH, for a sample with and without reduced latent heat effects, the natural evaporation rates are lower than water-only by 12.4% and 3.6% respectively. This is due to the porous material's thermal insulation effect, causing less heat to be transported from the ambient air, through the container and liquid water, to the evaporating surface. Most of the heat comes through the bottom pathway, as found from the previous FEA analysis and the correlation model. The inclusion of a porous material limits the heat transfer through the bottom pathway and causes the evaporation rate to drop.

The inclusion of intermediate water states with a reduced evaporation enthalpy of 50% increases the evaporation rate slightly when compared to sample evaporation without this effect. This is because there is less evaporative cooling effect on the air-porous evaporator interface, leading to the surface temperature to increase slightly and increase the surface vapor mole fraction. Due to energy conservation, there must be a cooling effect at the water-sample interface due to the difference in enthalpies between bulk water and intermediate water, which is commonly ignored in previous studies. This cools down the liquid water and draws additional environmental heat through the container's sidewalls. Since the total enthalpy difference needed to evaporate water depends on the initial state (bulk water) and the final state (water vapor) and all the heat ultimately comes from the ambient environment, the inclusion of intermediate water states with reduced latent heat inside of the porous evaporator doesn't change the total energy needed to evaporate water beyond its effects on the surface temperature of the evaporator. Furthermore, the total heat provided from the environment to evaporation is directly proportional to the evaporation rate through the latent heat of bulk water. Thus, Fig. 3B clearly illustrates that the total heat transferred from the environment is different for the sample evaporation and for water-only evaporation due to the difference in evaporation rates, invalidating the basic assumption of comparative natural evaporation experiments that there is equal heat input.

**Colder Surfaces Do Not Imply Faster Evaporation.** It has been hypothesized that natural evaporation rates from porous samples are faster than from water surfaces because the porous sample has a colder surface (*14*). The hypothesis is supported through simplified modeling arguments: colder interfaces lead to higher heat fluxes coming from the ambient environment due to larger temperature differences and lead to enhanced mass transfer coefficients due to larger temperature differences. This hypothesis was tested by modeling the evaporation from a porous sample as a function of its thermal resistance (Fig. 3C-D). The increasing thermal resistance of the porous material causes the surface temperature of the porous evaporator to decrease from 20.3 °C to 16.2 °C because less heat is transferred from the environment to the surface. However, the evaporation rate also decreases from 0.155 kg/m$^2$-hour to 0.121 kg/m$^2$-hour (Fig. 3D). This result can be understood through Eq. (1) described earlier and Fig. 3D. The evaporation rate is the product between the mass transfer coefficient, the inverse of $R_m''$, and the difference in vapor mole fraction between the surface of the evaporator and in the ambient air. Although the mass transfer coefficient increased by 18.9% with decreasing temperature, the saturated surface vapor molar concentration drops by 22.4%. The change in the saturated vapor molar concentration dominates over the increasing natural convection current, leading to overall decreasing evaporation rates with colder surface temperatures from porous materials.

The vapor kinetics depend on the vapor diffusivity in air, air's thermophysical properties, and the surface vapor mole fraction. Intermediate water doesn't couple to these properties beyond the surface temperature, which is shown to have minimal effects on predicted natural evaporation rates because the vapor kinetics doesn't change significantly in natural evaporation conditions. As a result, a reduced latent heat from intermediate water cannot lead to two-to-three-time enhancements in natural evaporation rates when compared to the water-only case. Furthermore, colder surfaces from porous materials lead to lower evaporation rates due to lower surface vapor content effects dominating over increased mass transfer coefficients.

**Systematic Error in Literature for Natural Evaporation Rates**. Comparisons are now made between existing literature data and the correlation model. Literature experimental sample size, relative humidity, and ambient temperature from dark evaporation experiments are inputted into the correlation model. The relative differences for evaporation rates between the reported literature results and the model predictions for water-only interfaces, normalized to the model predictions, are reported in Fig. 4. The model predictions assume water surfaces are at the same level as the container lips since no experiments had reported how much water recesses below the lip during dark evaporation. Due to the variety of experimental setups done in literature, such as open or closed environments and use of foam insulation on the surface, it cannot be expected that literature will match with this report's simulations exactly. However, it is still useful to compare the data to see if there are overarching trends in experiments. If evaporation rates from multiple different samples are reported in a work, we chose the sample that corresponds to the paper's reported high solar evaporation rates.

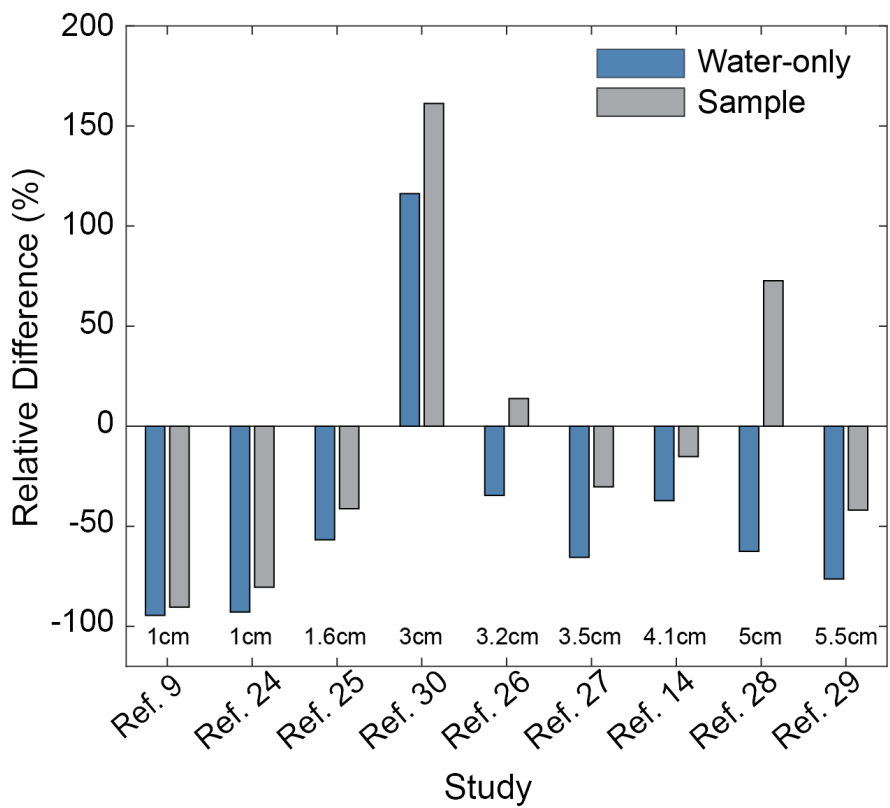

**Figure 4. Comparison with results in literature.** Relative difference of measured dark evaporation rates from water-only (blue) and samples (gray) to correlation model predictions (*9, 14, 24–30*). Sizes of samples are also displayed for clarity. Consistently lower measured evaporation from water-only surfaces suggests systematic error in previous works.

For almost all studies, the evaporation rate reported for water-only surface is systematically much lower than the predicted correlation data (*9, 14, 24–30*). Exact values for the data are shown in Table S1. Since the experimental results in this work measured evaporation rates within 10% of the model predictions for the RH conditions commonly used in experiments (30-50% RH), we believe that the predictions for dark evaporation experiments reported in literature should reasonably represent the expected evaporation rates if the water is levelled to the container lips (Table S2). For the porous sample data, the relative difference varies more but they are also typically lower than the model predictions as well. Some experiments in literature were conducted in open laboratory conditions, which could have forced convective currents and make the evaporation rate larger than the calculated values for natural evaporation rates. The relative differences are much larger than what can be explained as deviations in the reported ambient conditions and size of the evaporator based on the sensitivity analysis of the correlation model. Thus, the correlation models described earlier show precisely that porous evaporators do not have higher evaporation rates than water-only surfaces due to colder surface temperatures or potential reduced latent heat effects.

The fact that most reported water-only dark evaporation rates are lower than prediction indicates that water recessing below the containers' lip is the reason. Most porous materials also have dark evaporation rates comparable to or lower than prediction, suggesting they either remain aligned or recessed below the lip level. Thus, we have strong reason to believe that larger dark evaporation rates from porous samples are not due to reduced latent heat or other porous material effects, but simply reduced dark evaporation rates from water-only surfaces due to the water level recessing below the lip.

**Extension of Natural Convection Model to Solar Evaporation**. In solar evaporation experiments, a solar absorbing sample absorbs sunlight from the top surface and evaporates vapor at a higher temperature. Since the top surface is hotter than the ambient, natural convection causes the air current to move upwards. In the center region of the sample, a rising plume will form due to the unstable density stratification (*31*). Using this knowledge, we extended the correlation model to calculate the evaporation rates from solar experiments (see SI Note 4-6).

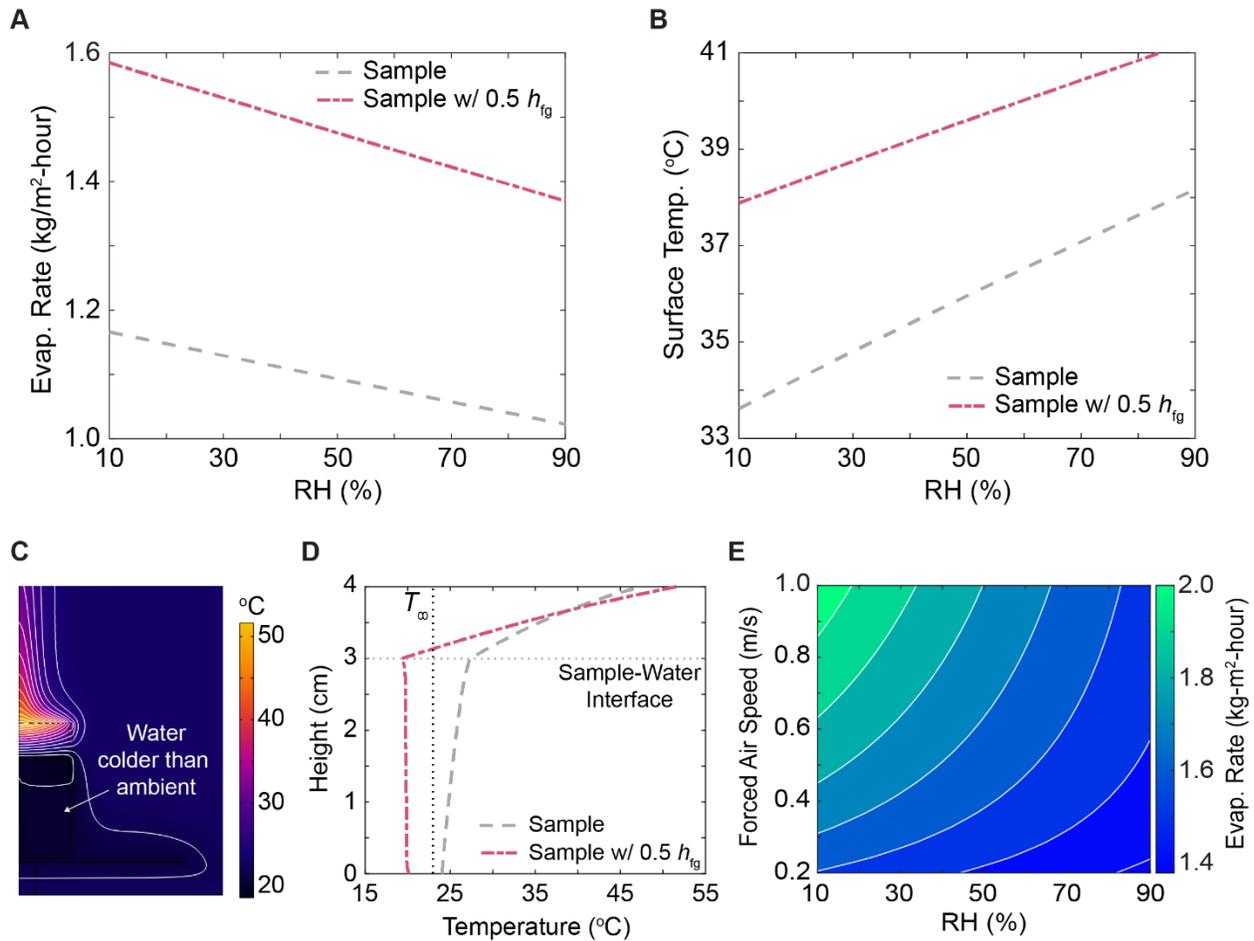

**Figure 5. Behavior under solar irradiation.** Predicted correlation model (**A**) evaporation rates and (**B**) surface temperatures during solar evaporation for a porous sample with and without latent heat reduction using the correlation model. The diameter is 3 cm, thermal resistance of 0.1 $m^2K/W$, and ambient temperature is 23 °C. (**C**) FEA simulation snapshot of the predicted temperature distributions, shown by the color gradient, for sample's solar evaporation rates with water inside the sample having half the latent heat. The ambient humidity is set to 30% RH with the other parameters set the same in (**A**)&(**B**). (**D**) FEA simulated temperature distribution along the height in the center of the evaporating setup for solar evaporation with and without intermediate water states. The horizontal dotted line shows the interface between the porous sample and water and the vertical dotted line shows the ambient temperature. Note the water temperature is below the ambient with the latent heat reduction assumption and above the ambient when no latent heat reduction (also in (**D**)). (**E**) Predicted evaporation rates under forced convection conditions for porous samples with intermediate water states using the correlation model.

In these correlation model calculations, we considered 2 cases: a porous sample base case in which there is no reduced latent heat and a porous sample with reduced latent heat as described in the previous section (SI Note 4-5). In all the cases, the incoming solar radiation of 1000 W/m$^2$ is assumed to be completely absorbed at the surface. We can see that the predicted evaporation rate for the sample with no reduced latent heat has evaporation rates ranging from 1.02 to 1.17 kg/m$^2$-hr depending on the ambient humidity. When there are intermediate water states with a reduced latent heat by 50%, the solar evaporation rates increase with a range from 1.37 to 1.58 kg/ m$^2$-hr (Fig. 5A). The predicted surface temperatures for samples with and without reduced latent heat ranges from 37.9 to 41.2 °C and 33.6 to 38.2 °C respectively (Fig. 5B). The surface temperature increases with intermediate water states because there is less evaporative cooling at the interface. These results have been validated with FEA simulations, which predict similar values and trends for both evaporation (Fig. S6A) and surface temperatures (Fig. S6B).

If the intermediate water states have large reduced latent heat effects, there is a large cooling effect at the sample-water interface from the enthalpy difference between bulk water and intermediate water. Fig. 5C illustrates the FEA steady-state temperature distribution after 2 hours of solar evaporation in simulation time. Inside of the container, the liquid water drops below the ambient temperature due to this cooling effect. When there are no intermediate water states, the liquid water is above the ambient temperature due to heat transfer from the hot evaporating surface (Fig. S6C). The temperature distributions for both cases along the centerline of the container from the liquid water bottom to the top of the sample evaporating surface is plotted in Fig. 5D. For sample evaporation without reduced latent heat, the surface centerline temperature is 47.4 °C and drops to 27.4 °C at the sample-water interface. Inside the liquid water, the temperature drops almost linearly from 27.4 °C to 24.0 °C due to the stable density stratification. In contrast for sample evaporation with reduced latent heat effects, the surface centerline temperature is higher at 51.6 °C and drops to 19.3 °C at the water-sample interface. Due to the unstable density stratification, the liquid water convectively mixes and has an almost isothermal temperature profile at about 19.8 °C. The bottom of the liquid water temperature increases to 20.1 °C due to the boundary layer between the bottom container wall and the liquid water. Using the correlation model, we also studied how the water temperature changes as a function of ambient relative humidity (Fig. S6). For samples with and without intermediate states, it was predicted that the mean water temperature ranges from 16.2 to 17.9 °C and 27.2 to 28.9 °C respectively as the ambient relative humidity increases from 10 to 90%.

From these pieces of evidence, it becomes clearer how intermediate water states could impact solar evaporation rates. The vapor kinetics increase due to less evaporative cooling effects at the evaporating surface and the surface heats up. Furthermore, the liquid water drops below ambient temperature due to the large cooling effect at the sample-water interface, causing it to draw in additional heat from the ambient environment through the container walls. However, a latent heat reduction of 50% does not increase the solar evaporation rates by a factor of 2 beyond the solar-thermal limit: the predicted evaporation rate from the correlation model at 30% RH is about 1.53 kg/m$^2$-hr, slightly above the thermal evaporation limit of 1.49 kg/m$^2$-hr if water is to evaporate thermally at the same surface temperature with 100% efficiency. In this case, the additional heat comes from the environment and transferred to the bulk water that is colder than the ambient. If all locations are above the ambient temperature, the sample can only lose heat to the environment. In these cases, superthermal evaporation rates are impossible even with the reduced latent heat hypothesis.

One final case considered is the effects of forced convection on solar evaporation. Solar evaporation experiments are typically conducted in open laboratory conditions because of physical constraints related to the size of the solar simulator, leading to the possibility of forced convection from outside sources such as air-conditioning. The Grashof number, $Gr$, governs the physics of air movement due to natural convection while the Reynolds number, $Re$, governs the intensity of air movement due to forced convection. The Grashof number is

$$Gr = \frac{gL_c^3 \rho_0 \Delta\rho}{\mu^2} \tag{4}$$

where $g$ is the gravitational acceleration, $L_c$ is the characteristic length of the evaporating surface, $\rho_0$ is the mean density in the boundary layer of evaporation, $\Delta\rho$ is the difference in densities between the evaporating surface and the ambient air, and $\mu$ is the dynamic viscosity. The Reynolds number is

$$Re = \frac{\rho_0 u L_c}{\mu} \tag{5}$$

where $u$ is the external airflow velocity due to forced convection. The ratio of the square root of the Grashof number to the Reynolds number $Re$ describes when these two terms are comparable in magnitude (*31*).

$$O(1) = \frac{Gr^{1/2}}{Re} \tag{6}$$

Using a sample with characteristic size of 3 cm, surface temperature of 45 °C, and ambient temperature of 23 °C, we can find that the forced convection velocity will need to be about 0.15 m/s. Forced convection could increase the evaporation rate due to a reduced mass transfer resistance for vapor as well as increased environmental heat input into the sub-ambient liquid water due to the cooling effect. Mixed convection, the combination of both natural and forced convection, is difficult to model accurately without the use of more sophisticated simulation methods, leading us to use correlations to model evaporation in the condition of forced convection in the crossflow (SI Note 6).

Using this correlation model, we tested to see if superthermal evaporation rates of 2 to 3 times the solar-thermal limit can be achieved using forced convection for the reduced latent heat hypothesis. The predicted evaporation rates as a function of both ambient humidity and external air velocity are illustrated in Fig. 5E. It can be seen evidently that as the external airflow increases in velocity, the evaporation rate increases as well, ranging from 1.38 to 2.05 kg/m²-hour as the external air velocity increases from 0.2 to 1.0 m/s and ambient humidity decreases from 90% to 10%. The increased evaporation rates can be attributed to lower mass transfer resistances from higher airflow velocities and higher environmental heat input into the cooled liquid water due to crossflow on the container sidewalls. The surface temperature decreases as well with increasing external airflow velocities due to larger evaporative cooling and heat loss to the environment (Fig. S8A). The liquid water temperature has a more complex behavior due to competing effects between increased environmental heating from higher air flow velocities, lower heat flux from the top evaporating surface due to lower surface temperatures, and varying sample-water interface cooling effects due to the changing evaporation rates (Fig. S8B). However, these high superthermal evaporation rates can only be achieved if there are high airflow velocities in the lab, very dry ambient conditions, and strong cooling effects, i.e., when the liquid water and/or the surface are below the ambient temperature. The cooling effect during superthermal evaporation experiments have not been reported in literature for 2D evaporators, leading us to conclude that reduced latent heat model due to an intermediate water state cannot explain superthermal evaporation rates.

**Potential Directions for Superthermal Evaporation.** Evidently, there is a need for both better experimental design and new directions to understand superthermal evaporation. The above discussion not only shows that the dark evaporation rates inference of reduced latent heat is prone to error, but also demonstrates that even if the hypothetically reduced latent heat was correct, it cannot lead to the doubling or tripling of the evaporation rates observed in many experiments. New mechanistic studies need to be rigorously conducted to understand how and why superthermal evaporation rates can occur. One potential theory that could explain this phenomenon is if water evaporates into an intermediate state, i.e. as water clusters, in the air instead of into single vapor (*9–11*). This could potentially explain superthermal evaporation rates because the enthalpy difference between bulk water and water clusters is smaller than that the difference between bulk water and water vapor, allowing energy to be conserved in solar evaporation experiments while allowing evaporation rates to exceed the solar thermal limit. Furthermore, evaporation as water clusters would increase the vapor concentration at the surface of the porous evaporator, leading to enhanced mass transfer kinetics as well. Although Zhao et al. mentioned this possibility, they introduced the reduced latent heat mechanism, i.e., a thermal evaporation picture (*9*).  However, Tu et al. showed via experiments that heating alone by electrical current nor by light cannot lead to superthermal evaporation and introduced the photomolecular evaporation mechanism: light can directly cleave off water molecular clusters (*10*). Subsequent studies lent more support for the photomolecular picture (*11, 17, 18*), but significant more work is needed to test this picture rigorously.

## DISCUSSION

Reduced latent heat due to intermediate states of water inside of interfacial solar evaporators is a common hypothesis to explain superthermal evaporation rates. To quantify the reduced latent heat, comparison of the natural evaporation rates between a water-only surface and a porous surface in apparent identical conditions is used to justify that water in porous materials has a reduced latent heat.  Results presented above strongly indicate that such a comparison is not sound for one major reason. Dark evaporation rates from water are very sensitive to the height of the water level relative to the mouth of the container: a recessed height of a few mm can decrease the evaporation rate by about 50%. Our scaling analysis also suggests that higher natural evaporation rates from increased internal surface areas due to microporosity of 2D surface evaporators is unlikely because the dominating mass transfer resistance is in the air side boundary layer. We advise against using these experiments to draw conclusions in changes in latent heat of water because of the sensitivity of these measurements and the lack of standardization has led to problematic conclusions on evaporation rate improvements from porous materials.

Our models and simulations illustrate that the reduced latent heat hypothesis does not lead to higher natural evaporation rates, as previously assumed. What was neglected in such an argument is that when bulk water enters the porous material, a cooling effect will happen at the sample-water interface due to the difference in water enthalpy. Thus, similar amounts of heat equaling to the bulk latent heat of water must still be provided through the environment. Furthermore, for natural convection cases, the air side mass transfer resistance cannot change significantly, leading to only small changes in predicted dark evaporation rates. For solar evaporation, our results show that reduced latent heat leads to a small increase in evaporation rates in natural convection conditions, but never the high superthermal rates reported in literature. With forced convection, evaporation rates could exceed the thermal limit in extremely dry ambient conditions and high convective airflows due to the lower mass transfer resistance and higher environmental heat input into the liquid water. In all cases for solar evaporation with reduced latent heat from intermediate water states inside of the porous evaporator, the liquid water should achieve sub-ambient temperatures due to the large cooling effect, which has never been reported in literature. In no conditions can we simulate 2-3 times evaporation rate due to latent heat reduction. We emphasize that a simple reduced latent heat picture is erroneous, which lends more credence to alternative explanations such as the photomolecular mechanism (*10, 11, 17*).

**METHODS**

**COMSOL FEA Details**. We have modeled natural evaporation in COMSOL using FEA in transient simulations similar to what is conducted in laboratory experiments. We have fully simulated liquid water inside a plastic container in contact with open air using a 2D axis-symmetric setup. The container has a thickness of 2 mm, height of 4 cm, and varying diameters from 1 to 5 cm. Water inside container is also fully simulated. Open air is fully simulated around the container, creating a hemispherical simulation domain with a radius of 1 m. An image of the generated mesh for this domain is illustrated in Fig. S9. The general governing equations for mass, momentum, energy, and vapor transport are

$$\frac{\partial \rho}{\partial t} + \nabla \cdot (\rho \vec{u}) = 0 \tag{7}$$

$$\rho \frac{\partial \vec{u}}{\partial t} + \rho (\vec{u} \cdot \nabla)\vec{u} = -\nabla p + \nabla \cdot \left(\mu(\nabla \vec{u} + (\nabla \vec{u})^T) - \frac{2}{3}\mu(\nabla \cdot \vec{u})I\right) + \rho \vec{g} \tag{8}$$

$$\rho C_p \left(\frac{\partial T}{\partial t} + \vec{u} \cdot \nabla T\right) = \nabla \cdot (k \nabla T) \tag{9}$$

$$\frac{\partial c_v}{\partial t} + \nabla \cdot (c_v \vec{u}) = \nabla \cdot (D_v \nabla c_v) \tag{10}$$

where $\rho$ is the material density, $u$ is the fluid velocity, $t$ is time, $p$ is the gauge pressure, $\mu$ is the fluid viscosity, $g$ is gravity, $C_p$ is the isobaric specific heat capacity, $T$ is the temperature, $k$ is the thermal conductivity, $c_v$ is the mole fraction of vapor, and $D_v$ is the vapor diffusivity in excess of air. The vapor content effects on the air-vapor mixture thermodynamic properties are ignored to expedite calculations due to the relatively low saturation vapor pressures expected. This will allow us to assume that the air's thermophysical properties are the same as dry air. The fluid properties are temperature dependent, leading to natural convective forces being induced using the weakly compressible model. An open boundary condition is set for the top of the hemispherical dome with a gauge pressure set to 0 Pa, prescribed ambient temperature at 23 °C, and prescribed ambient humidity. The water container is set on a solid substrate kept at ambient temperature. The influence of setting the outer wall of the container bottom to the ambient temperature in natural convection cases is analyzed using the correlation model in SI Note 7. Using the correlation model, it was found that for natural evaporation of water-only interfaces, the inclusion of the effective heat transfer resistance of the scale's weighing pan changed the predicted evaporation rate by about 5% averaged across RH. This insensitivity is because (1) the heat transfer through the container sides is a parallel channel for heat transfer and is larger than the heat flux from the bottom, (2) the vapor kinetics doesn't change significantly during natural evaporation due to the rather small changes in surface temperature, and (3) the low evaporative cooling fluxes make the temperature difference between the water and ambient small.

At the interface between solids and fluids, the no-slip boundary condition is imposed. At the interface between the liquid and air, it is assumed that the vapor concentration is always at the saturated vapor pressure for its given temperature. Significant enhancement of the mesh is made in the region close to the evaporating surface as well as in the boundary layers due to these conditions. The initial conditions for the entire domain were set to be isothermal at the ambient temperature and the vapor pressure set at the ambient value. At the air and water interface, an additional boundary heat flux is added to include the effects of evaporative cooling and radiation exchange with the ambient air.

$$q_s''(\vec{r}, t) = -h_{fg}(T)M_v \dot{m}''(\vec{r}, t) + \sigma \epsilon_w \left(T_\infty^4 - T^4(\vec{r}, t)\right) \tag{11}$$

where $\vec{r}$ is the local coordinates in the simulation, $t$ is the simulation time, $\sigma$ is the Stefan-Boltzmann constant, and $\epsilon_w$ is the emissivity of water. As a result, the boundary heat flux is calculated locally at the air-liquid interface. The local evaporation flux, $\dot{m}''(\vec{r}, t)$, comes directly from solving Eq. (10) at each timestep and the local temperatures are calculated from the solving the heat equations. This leads to an energy relationship between the evaporative heat flux, radiative heat transfer, and thermal conductance from the air and from the water at the interface. At steady state, these heat fluxes will balance to zero.

$$0 = q_s''(\vec{r}, t) + k_{air}\frac{dT_{air}}{dz} - k_{water}\frac{dT_{water}}{dz} \tag{12}$$

The difference in signs for the thermal conductance of air and water is due to directionality of heat transport to the interface and these are solved for directly from the energy transport equations. The transient simulations were solved for 2 hours of simulation time. The last half hour's data were used as steady-state data to conduct analysis for this work.

For solar evaporation of porous samples with and without reduced latent heat effects, three additional changes are made to the simulation domain. The aluminum pan from the weighing scale is directly simulated underneath the evaporating container with the same geometric features as described in SI Note 7. This feature was added to relax the bottom ambient temperature assumption from natural evaporation simulations due to the larger evaporation fluxes expected in solar evaporation. A zoomed in image of the mesh is illustrated in Fig. S10. The simulation domain of the air was also reduced to 0.6 m in radius to reduce the computation time of the simulations. Only heat conduction is assumed to happen in the porous evaporator. Since solar-thermal evaporation rates are on the order of 1 kg/m²-hr, the net mass transport of liquid water through the porous evaporator is on the order of 0.3 µm/s, justifying the above assumption.

The sample's thermophysical properties are set to 1 cm thick, a thermal conductivity of 0.1 W/m-K, a specific heat capacity of 2000 J/kg-K, and a density of 1200 kg/m³.

To account for solar absorption, it is assumed that the sample has 100% absorptance across the solar spectrum and has a small optical penetration depth, leading to an additional term in Eq. (11) to account for 1 sun intensity $q''_{sun}$.

$$q''_s(\vec{r}) = -h_{fg}(T)M_v \dot{m}''(\vec{r},t) - \sigma\epsilon_w(T^4(\vec{r},t) - T^4_\infty) + q''_{sun} \qquad (13)$$

For simulations with latent heat reduced by χ (we used χ=0.5 or 50% for all calculations) due to intermediate water states in porous materials, an additional pre-factor was included for the evaporation heat flux.

$$q''_s(\vec{r}) = -\chi h_{fg}(T)M_v \dot{m}''(\vec{r},t) - \sigma\epsilon_w(T^4(\vec{r},t) - T^4_\infty) + q''_{sun} \qquad (14)$$

To include the cooling effect at the bottom interface between the solid and liquid water, the enthalpy difference is averaged over the top surface area and evenly distributed over the sample-water surface area. This calculation assumes that the cooling effect is based on the instantaneous evaporation rate at that time $t$. There would be a time lag between these two due to the thermal capacitance of the porous evaporator and the liquid water flowrates inside of the material. Since only steady-state results are analyzed, we expect the transient effects to not have an impact on the result. This leads to the boundary heat flux at the sample-water interface, $q''_{s,w}$, to be

$$q''_{s,w} = -\frac{(1-\chi)\int \dot{m}_{evap}(\vec{r},t)h_{fg}(T)d\vec{r}}{A_c} \qquad (15)$$

where the top integral is over the sample-air interface area $A_c$ and the $(1-\chi)$ pre-factor accounts for the enthalpy difference between bulk water and intermediate water.

**Experimental Details**. All evaporation experiments are conducted in a sealed stainless-steel chamber with characteristic dimensions of 0.6 m in each axis. A schematic of the setup is illustrated in Fig. S11A. An inlet allows controlled humidity air flow into the chamber at a fixed flow rate while an outlet is connected to the lab to maintain atmospheric pressure inside of the chamber. The dry air has a humidity of about 10% and its humidity is changed by flowing it through containers of saturated salt solutions. The salt solutions used are $MgCl_2$, $K_2CO_3$, and NaCl. All water used is Type 1 Ultrapure water.

For the experiments, we made a 3D printed container from polylactic acid (Fig. S11B). The evaporating surface has a radius of 1.5 cm and a height of 4 cm. The connected reservoir has a cross-sectional area about 4.5 times larger than the evaporating surface. The reservoir is used to ensure that the evaporating height is steady during the entire experiment. The reservoir is sealed using a snap on lid and parafilm. The top of the reservoir lid has a 0.2 mm diameter hole to maintain ambient pressure inside of the reservoir. Initially, the chamber's air is replaced with controlled humid air during an equilibration step using high flowrates for 3 hours. During this state, the evaporating surface is covered with a lid controlled by a stepper motor to prevent evaporation. The flow rate is then reduced to 3.5 L/min for at least 30 minutes before the weight loss is recorded. Once the equilibration step is completed, the stepper motor lifts the cap above the evaporating surface by 10 cm and the mass loss is recorded using a digital scale. All evaporation rates are measured for 2 hours, but only the last hour is used for data analysis. Tests have been conducted to find that the used air flowrate has minimal impact on the measured evaporation rate (Fig. S12). The humidity and temperatures inside the chamber are constantly monitored using Honeywell HIH8000 sensors. Data collection and the stepper motors are controlled through MATLAB and Arduino setups.

**Correlation Model Details**

A correlation model is developed to describe evaporation rates from water-only and porous evaporators inside of a cylindrical container under natural convection, solar input with natural convection, and solar input with forced convection into a large ambient reservoir of air at a prescribed temperature and relative humidity. The full heat and mass transfer models are described in SI Notes 2-7. All the heat and mass transfer resistances considered are inputted into a MATLAB program. Air and water's thermophysical properties are described using the film temperature between the solid interface and the bulk fluid it is in contact with. Energy conservation equations are then applied at each of the temperature nodes described in Fig. S2 and Fig. S5. The node temperatures and coefficients are iterated until a self-consistent solution is found. For natural evaporation, the vapor content effects on air's thermophysical properties are ignored due to its low concentrations. For solar evaporation calculations, the vapor content effects are incorporated when calculating air's thermophysical properties.

**Acknowledgements**

The authors would like to acknowledge Dr. Wenhui Tang for her feedback on the figures.

**Funding**: This project is supported through the Abdul Latif Jameel Water and Food Systems Lab (J-WAFS), UM6P & MIT Research Program (UMRP), and MIT Bose Award. J.H.Z. acknowledges support from the Abdul Latif Jameel Water and Food Systems Lab Graduate Fellowship and the MATHWORKS Graduate Fellowship.

**Authors contribution:** Conceptualization: J.H.Z, R.M., and G.C. Experimentation: J.H.Z. Computations: J.H.Z. Analysis: J.H.Z. Resources: G.C. Writing: J.H.Z. Review and edit: G.C., R.M., and G.L. Visualization: J.H.Z. Supervision: G.C.

**Competing interests:** The authors declare that they have no competing interests.

**Data and materials availability:** All details on analysis and experimentations are fully detailed in the manuscript and/or Supplementary Information.

Supplementary Information for

# Reexamination of evaporation from horizontal surfaces with implications for solar interfacial evaporation experiments

James H. Zhang[1], Rohith Mittapally[1], Guangxin Lv[1], Gang Chen[1*]

[1]Mechanical Engineering Department, Massachusetts Institute of Technology,

Cambridge, MA 02139

*Corresponding author: gchen2@mit.edu

**Supplementary Information Note 1**

**Estimation of temperature difference to cause convective mixing in water**

Due to the cold evaporating top surface and relatively warmer temperature of water on the bottom, the unstable density stratification in the liquid water causes natural convection currents to form. Laminar flow of water inside the container will cause mixing and lead to a more isothermal temperature distribution.

The water in the container can be treated as in an enclosure heated on the bottom and sidewalls and cooled on the top surface from evaporation. A critical Rayleigh number commonly cited for convection to take place for a free top surface and rigid bottom surface is (*32*)

$$Ra_H \geq 1100 \tag{S1.1}$$

Using the thermophysical properties of water, we can solve for the $\Delta T$ that will lead to convection.

$$\Delta T = \frac{1100\alpha\nu}{g\beta H^3} = 0.0012\ K \tag{S1.2}$$

where $g$ is gravity, $\beta$ is the thermal expansion of water, $H$ is the height between the hot bottom and cold top, $\alpha$ is the thermal diffusivity of water, and $\nu$ is the kinematic viscosity of water. Plugging in values, we can find the critical $\Delta T$ is 1.2 mK.

**Supplementary Information Note 2**

**Correlation Model to Predict Water-Only Natural Evaporation**

We will construct a simplified model to estimate the evaporation rates from a pure water surface. The full heat and mass transfer diagram is illustrated in Fig. S2. To model the top air side heat transfer, we will use Kadambi and Drake's correlation for circular cold plate in hot environment (*16, 33*). We will assume that the heat and mass transfer analogies hold, so that the same correlation can model both the Sherwood number and the Nusselt number.

$$Sh_D = 0.82 Gr_D^{0.2} Sc^{0.234} \tag{S2.1}$$

$$Nu_D = 0.82 Gr_D^{0.2} Pr^{0.234} \tag{S2.2}$$

$D$ is the diameter of the evaporating surface, $Gr$ is the Grashof number, $Sc$ is the Schmidt number, and $Pr$ is the Prandtl number. Since squares and circles have the same area to perimeter ratio, the same values can be calculated for squares but instead using its side-length. The Sherwood and Nusselt numbers can be related to the air side heat and mass transfer coefficients through

$$g_m = \frac{Sh_D D_v}{D} \tag{S2.3}$$

$$h_{air} = \frac{Nu_D k_{air}}{D} \tag{S2.4}$$

where $D_v$ is the diffusion coefficient of water vapor in excess of air and $k_{air}$ is the thermal conductivity of air. The heat and mass transfer coefficients can be used to find the air side heat transfer and evaporation rate through Eqs. (S2.5-S2.7).

$$\dot{m} = g_m A_c C_g \left( c_{v,s}(T_s) - (RH) c_{v,s}(T_\infty) \right) \tag{S2.5}$$

$$q_{evap} = \dot{m} h_{fg}(T_s) \tag{S2.6}$$

$$q_{air} = h_{air} A_c (T_\infty - T_s) \tag{S2.7}$$

where $A_c$ is the cross-sectional area, $C_g$ is the molar density of dry air, $c_{v,s}$ is the saturated vapor mole fraction at a certain temperature, $RH$ is the ambient relative humidity, $h_{fg}$ is the latent heat of evaporation, $T_s$ is the evaporating surface temperature, and $T_\infty$ is the ambient temperature. Radiation exchange between the top surface and the ambient is modeled using the typical heat transfer between a surface and a large gray body reservoir.

$$q_{rad} = h_{rad} A_c (T_\infty - T_s) \tag{S2.8}$$

where $h_{rad}$ is

$$h_{rad} = \sigma \epsilon_w (T_\infty^2 + T_s^2)(T_\infty + T_s) \tag{S2.9}$$

$\sigma$ is the Stefan-Boltzmann constant and $\epsilon_w$ is the blackbody emissivity of water. Heat transfer through the water pathway is modeled as a resistance network from the outside ambient air to the evaporating top surface. This leads to the general heat transfer equation to become

$$q_{water} = \frac{(T_\infty - T_s)}{R_{water}} \tag{S2.10}$$

where $R_{water}$ is the total resistance of the bottom pathway from the ambient environment to the evaporating surface. We assume that the bulk water is isothermal with its surface temperature due to convective flows, leading to temperature differences only occurring at the boundary layers near the container walls. The heat transfer resistance from the ambient environment through the water to the surface comes from two parallel pathways: the bottom of the container and the side of the container.

$$\frac{1}{R_{water}} = \frac{1}{R_{bot}} + \frac{1}{R_{side}} \tag{S2.11}$$

The convective flow boundary layers of water on the sidewalls and the bottom container wall should be coupled due to the Rayleigh-Bernard convection cell, however, we will calculate the heat transfer through each boundary layer separately due to the lack of heat transfer empirical correlations for the prescribed flow conditions. The resistance through the bottom is a series resistance of the aluminum scale pan (SI Note 7), acrylic container wall, and natural convection from the warmer acrylic container bottom into the colder water.

$$R_{bot} = R_p + \frac{1}{A_c}\left(\frac{t_{cont}}{k_{acry}} + \frac{1}{h_{bot}}\right) \tag{S2.12}$$

where $R_p$ is the effective heat transfer resistance from the aluminum scale pan found in SI Note 7. The convective heat transfer coefficient from the bottom of the acrylic container to the water was approximated using Raithby and Hollands' correlation for a hot plate in cold fluid environment (*16, 34*).

$$h_{bot} = \frac{Nu_{L^*,corr} k_{water}}{L^*} \tag{S2.13}$$

$$Nu_{L^*,bot} = \frac{0.56 Ra_{L^*,water}^{1/4}}{(1 + (0.492/Pr_{water})^{9/16})^{4/9}} \tag{S2.14}$$

Due to the low Nusselt numbers expected from the small temperature differences, the Nusselt number in Eq. (S2.14) is corrected using the following equation (*16*).

$$Nu_{L^*,corr} = \frac{1.4}{\ln\left(1 + \frac{1.4}{Nu_{L^*,bot}}\right)} \tag{S2.15}$$

The characteristic number used for Raithby and Hollands correlation is based on the area to perimeter ratio of the bottom surface (*16, 35*).

$$L^* = \frac{A_c}{P} = \frac{D}{4} \tag{S2.16}$$

The heat transfer through the sidewalls has three heat transfer resistances in series: air side convection, acrylic sidewall, and water convection.

$$R_{side} = \frac{1}{A_p}\left(\frac{t_{acry}}{k_{acry}} + \frac{1}{h_{side,air}} + \frac{1}{h_{side,water}}\right) \tag{S2.17}$$

Both the air and water natural convection correlations on the sidewalls were evaluated using Churchill and Chu's correlation for a vertical wall in natural convective conditions for the respective fluid thermophysical properties (*36*).

$$Nu_{H,side} = 0.68 + \frac{0.67 Ra_H^{1/4}}{(1 + (0.492/Pr)^{9/16})^{4/9}} \tag{S2.18}$$

$$h_{side,air} = \frac{Nu_{H,side} k_{air}}{H} \tag{S2.19}$$

$$h_{side,water} = \frac{Nu_{H,side} k_{water}}{H} \tag{S2.20}$$

where $Ra$ is the Rayleigh number describing the natural convective flow. For these calculations, we made the thin wall approximation to neglect the changing thickness due to the curvature of the container.

$$t_{acry} \ll D \tag{S2.21}$$

All fluid properties, non-dimensional numbers, convective heat transfer coefficients, and convective mass transfer coefficients are evaluated at the mean temperature between the wall temperature and the bulk fluid temperature it is in contact with.

$$T_{film} = \frac{(T_{fluid} + T_{wall})}{2} \tag{S2.22}$$

As a result, the temperature of the evaporating surface, $T_s$, can be found through energy balance on the surface temperature node.

$$0 = -q_{evap} + q_{air} + q_{rad} + q_{water} \tag{S2.23}$$

This equation can be used to solve for the unknown surface temperature for a given geometry and two boundary conditions of the ambient RH and ambient temperature using nonlinear zero solvers. The evaporation rate can then be found by using Eq. (S2.5) for the calculated surface temperature.

**Supplementary Information Note 3**

**Correlation Model to Predict Porous Sample Evaporation**

The addition of a porous material will have two effects on the overall heat transfer in the model: additional heat transfer resistances through the bottom waterside pathway and the inclusion of intermediate water states. The full heat and mass transfer diagram is shown in Fig S5 with all temperature and vapor concentration node labels. At the top interface, Eq. (S2.23) and $q_{water}$ from Eq. (S2.10) must be modified to account for the thermal insulation effect and reduced latent heat effects.

First, if there are no reduced latent heat effects as seen for the base case porous sample evaporation, two additional heat transfer resistances that are in series with the original heat transfer resistance through the water need to be added, leading to $q_{water}$ in Eq. (S2.10) being replaced with $q_{samp}$.

$$q_{samp} = \frac{(T_\infty - T_s)}{R_{water} + R_{s,w} + \frac{t_{samp}}{k_{samp} A_c}} \tag{S3.1}$$

The addition of the porous materials will create another boundary layer at the surface where water and the sample are in contact, leading to the additional thermal resistance $R_{s,w}$.

$$R_{s,w} = \frac{1}{h_{s,w} A_c} \tag{S3.2}$$

At the sample-water interface with temperature $T_{s,w}$, the water density stratification is not stable due to the bulk water, $T_w$, having a warmer temperature. As a result, $h_{s,w}$ can be described using Raithby and Hollands' correlation with the corresponding film properties in Eq. (S2.14-2.16). We will ignore the detailed flow of water inside of the porous evaporator and lump it into the sample's effective thermal conductivity, leading to the third thermal resistance term in Eq. (S3.1). This is because the average net liquid water flow velocity is very small due to the low evaporation rates expected from natural evaporation. Since the evaporating surface is no longer a water-only interface, the surface temperature $T_s$ is not equal to the bulk water temperature $T_w$. All other equations for describing $R_{water}, q_{rad}, q_{evap},$ and $q_{air}$ in SI Note 2 still hold.

Eq. (S3.1) is only true if there is no cooling effect at the sample-water interface due to the enthalpy difference between bulk water and intermediate water states from reduced latent heat. If reduced latent heat effects are to be included, energy balance needs to be applied at the surface temperature node $T_s$ and the sample-water interface temperature node $T_{s,w}$. At the surface temperature node, Eq. (S2.23) becomes

$$0 = -\chi q_{evap} + q_{air} + q_{rad} + \frac{k_{samp} A_c}{t_{samp}} (T_{s,w} - T_s) \tag{S3.3}$$

and at the sample-water interface

$$0 = -\frac{k_{samp} A_c}{t_{samp}} (T_{s,w} - T_s) + \frac{(T_\infty - T_{s,w})}{R_{s,w} + R_{water}} - (1 - \chi) q_{evap} \tag{S3.4}$$

To account for the reduced latent heat of evaporation of intermediate water $\chi$, a cooling effect occurs at the sample and water interface due to the enthalpy difference between bulk water and intermediate water. For simplicity of this analysis, we set the enthalpy of the intermediate state to be halfway between bulk liquid and water, leading to a latent heat of evaporation reduction of 50% ($\chi=0.5$). As a result, we distribute the evaporation heat flux, $q_{evap}$, equally to the top and bottom side of the sample. We then solve these equations to find the evaporation rate of porous evaporators. This analysis assumes that the surface of the interfacial evaporator is fully wetted and hydrodynamically smooth. Hydrodynamically smooth refers to the surface roughness characteristic size to be small when compared to the momentum boundary layers.

**Supplementary Information Note 4**

**Correlation Model to Predict Porous Sample Solar Evaporation without Reduced Latent Heat**

In solar evaporation, the top side of the porous evaporator is now a hot surface in a cold environment. As a result, we will now use Raithby and Holland's correlation to estimate the top air side heat transfer evaporation using the heat and mass transfer analogy again (*16, 34*).

$$Nu_{L^*,air} = \frac{0.56 Ra_{L^*}^{1/4}}{(1+(0.492/Pr)^{9/16})^{4/9}} \quad (S4.1)$$

$$Sh_{L^*,evap} = \frac{0.56 Gr_{L^*}^{1/4} Sc^{1/4}}{(1+(0.492/Sc)^{9/16})^{4/9}} \quad (S4.2)$$

Due to the low Rayleigh numbers expected, the $Nu_{L^*,air}$ is corrected again using Eq. (S2.15). $Sh_{L^*,air}$ will be corrected using the same equation.

$$Sh_{L^*,corr} = \frac{1.4}{\ln\left(1 + \frac{1.4}{Sh_{L^*,evap}}\right)} \quad (S4.3)$$

Thus, the heat and mass transfer on the top side of the porous evaporator, $q_{air}$, $\dot{m}_{evap}$, and $q_{evap}$ can be described by using the corrected Nusselt numbers and Sherwood numbers in Eqs. (S4.1-4.3).

Since there is no convection in water due to the stable density stratification, we can treat heat transfer in water as due to conduction only. In this case, the Biot number in the radial direction, $Bi = h_{eff} R/k_w$, is ~0.25. Hence, we can approximate water as a fin with temperature variation only along the height z-direction, described by the fin equation

$$\frac{d^2 T_w}{dz^2} = \frac{h_{eff} P}{k_w A_c}(T_w - T_\infty) \quad (S4.4)$$

where we define the base of the fin as the interface between the sample and water and the z-axis to point downwards. The effective heat transfer coefficient, $h_{eff}$, includes the resistance of the container wall as well as the natural convection coefficient.

$$h_{eff} = \frac{1}{\frac{t_{cont}}{k_{cont}} + \frac{1}{h_{side}}} \quad (S4.5)$$

Using a change of variables

$$\theta = T_w - T_\infty \quad (S4.6)$$

and integrating Eq. (S4.4) yields the general fin equation

$$\theta = c_1 e^{z\sqrt{\frac{h_{eff} P}{k_w A_c}}} + c_2 e^{-z\sqrt{\frac{h_{eff} P}{k_w A_c}}} \quad (S4.7)$$

The heat flux and temperature at the interfaces between the sample-water and sample-container bottom must be continuous, yielding four boundary conditions. At the sample-water interface

$$-k_{water}\frac{d\theta}{dz}\bigg|_{z=0} = \frac{k_{samp}}{t_{samp}}(T_s - T_{s,w}) \tag{S4.8}$$

$$\theta(0) + T_\infty = T_{s,w} \tag{S4.9}$$

At the bottom-water interface

$$-A_c k_{water}\frac{d\theta}{dz}\bigg|_{z=H} = q_{bot} \tag{S4.10}$$

$$\theta(H) + T_\infty = T_{bot} \tag{S4.11}$$

The heat transfer through the bottom of the container is

$$q_{bot} = \frac{(T_{bot} - T_\infty)}{R_p + \frac{t_{cont}}{k_{cont}A_c}} \tag{S4.12}$$

where $T_{bot}$ is the temperature of the container bottom wall in contact with the water and $R_p$ is the effective aluminum scale pan resistance found in SI Note 7. $h_{side}$ in Eq. (S4.5) depends on the outer container wall temperature and there is a temperature distribution inside of the water, leading to the need to approximate $h_{side}$ for this analysis. Using a 1D heat transfer network, we approximated the outer container wall temperature node $T_o$ and $h_{side}$ by doing an energy balance on the outer container wall temperature node. In this analysis, we assume that the inside wall temperature of the acrylic container is isothermal with the water: $T_i(z) = T_w(z)$.

$$T_{w,ave} = \frac{\int_0^H T_w(z)dz}{H} \tag{S4.13}$$

$$\frac{k_{cont}(T_{w,ave} - T_o)}{t_{cont}} = h_{side}(T_o - T_\infty) \tag{S4.14}$$

Eq. (S2.23) is modified to account for the solar absorption energy for energy balance at the sample-air interface.

$$0 = q_{sun} - q_{evap} + q_{air} + q_{rad} - \frac{k_{samp}A_c}{t_{samp}}(T_s - T_{s,w}) \tag{S4.15}$$

where $q_{sun}$ is set to one-sun intensity.

$$q_{sun} = 1000A_c \tag{S4.16}$$

$q_{air}$ and $q_{rad}$ have different signs because it is defined in Eq. (S2.7) and Eq. (S2.8) as the difference between the ambient temperature and the surface temperature. Since the surface temperature is hotter than ambient, these terms will become negative and sources of heat loss in Eq. (S4.15). The evaporation rate can then be solved by numerically solving for each of the temperature nodes and the unknown fin coefficients in Eq. (S4.7) by using the boundary conditions Eqs. (S4.8-4.11).

**Supplementary Information Note 5**

**Correlation Model to Predict Porous Sample Solar Evaporation with Reduced Latent Heat**

Since the topside is hot, the heat transfer and mass transfer equations from Eqs. (S4.1-4.3) still hold for describing $q_{air}$ and $q_{evap}$. However, the bottom interface of the porous evaporator will no longer be hot because of the mixing cooling effect when bulk water becomes an intermediate water state from reduced latent heat effects. As a result, the water temperature profile is no longer stratified in a stable density configuration and leads to natural convective mixing. This causes Eqs. (S4.4-4.10) to no longer hold. The energy balance on the top interface in Eq. (S4.14) becomes

$$0 = q_{sun} - \chi q_{evap} + q_{air} + q_{rad} - \frac{k_{samp} A_c}{t_{samp}}(T_s - T_{s,w}) \qquad (S5.1)$$

where the latent heat of evaporation is reduced by half ($\chi$=0.5). At the bottom interface with temperature $T_{s,w}$, the energy balance becomes the same as Eq. (S3.4).

$$0 = \frac{k_{samp} A_c}{t_{samp}}(T_s - T_{s,w}) + \frac{(T_\infty - T_{s,w})}{R_{s,w} + R_{water}} - (1-\chi)q_{evap} \qquad (S5.2)$$

where $T_w$ is the bulk water temperature. Eq. (S3.2) and Eqs. (S2.11-2.20) still hold to describe the heat transfer resistances from the boundary layer at the sample-water interface $R_{s,w}$ and the parallel heat transfer network from the ambient air into bulk water $R_{water}$.

**Supplementary Information 6**

**Correlation Model to Predict Porous Sample Solar Evaporation Under Forced Convection**

For forced convection, Eqs. (S4.1-4.2) and (S2.18-2.19) will be modified to forced convection correlations. Since the airspeeds are likely to be low, we will be in the laminar regime. The heat and mass transfer above the evaporating surface will be approximated using crossflow over a flat plate correlation (*16*).

$$Nu_{D,air} = 0.664 Re_D^{1/2} Pr^{1/3} \tag{S6.1}$$

$$Sh_{D,evap} = 0.664 Re_D^{1/2} Sc^{1/3} \tag{S6.2}$$

The sidewall heat transfer coefficient from air to the container wall will use the crossflow correlation for low Reynolds number flows against cylinders (*31, 37*).

$$Nu_{D,side} = 0.3 + \frac{0.62 Re_D^{1/2} Pr^{1/3}}{(1 + (0.42/Pr)^{2/3})^{1/4}} \tag{S6.3}$$

$$h_{side,air} = \frac{Nu_{D,side} k_{air}}{D} \tag{S6.4}$$

**Supplementary Information 7**

**Scale Pan Effective Thermal Resistance**

Although in the FEA simulation, we assumed that the bottom surface is at ambient temperature, most experiments have the container bottom sitting on the surface of a weighing scale, and hence questions can be raised on how accurate is the assumption of constant ambient temperature at the bottom surface. In this case, ambient heat is transferred to the bottom via air convection onto the scale pan and conduction along the pan.

Due to the very laminar features, we will assume a constant natural convection coefficient of air of 1 W/m²-K on the aluminum pan for this analysis. Heat transfer through the bottom pathway goes through natural convection of the aluminum pan on the scale and into the water bottom. The scale's aluminum pan can be modeled as an annular fin.

$$k_p t_p \frac{\partial}{\partial r}\left(r \frac{\partial T}{\partial r}\right) - 2h(T - T_\infty) = 0 \tag{S7.1}$$

The resulting equation then becomes

$$r \frac{\partial^2 T}{\partial r^2} + \frac{\partial T}{\partial r} - \frac{2hr}{k_p t_p}(T - T_\infty) = 0 \tag{S7.2}$$

Using a change of variable

$$\theta = \frac{T - T_\infty}{T_b - T_\infty} \tag{S7.3}$$

leads to

$$r \frac{\partial^2 \theta}{\partial r^2} + \frac{\partial \theta}{\partial r} - \frac{2hr}{k_p t_p}\theta = 0 \tag{S7.4}$$

The boundary conditions for this equation are

$$\theta(r = r_c) = 1 \tag{S7.4}$$

and

$$-k_p \frac{d\theta}{dr}\Big|_{r=r_p} = h\theta(r = r_p) \tag{S7.5}$$

where $r_c$ is the radius of the container and $r_p$ is the radius of the aluminum pan. Assuming all the heat absorbed by the scale is transferred to the water container, the total heat from the annular aluminum plate fin to the base of the container is equal to

$$q_p^* = -k_p(2\pi r_c t_p)\frac{d\theta}{dr}\Big|_{r=r_c} \tag{S7.6}$$

where $r_c$ is the radius of the container $t_p$ is the thickness of the aluminum pan. We can then recognize that this gives us an effective thermal resistance to describe heat transfer from the ambient air, through the aluminum pan, and into the bottom of the water container. First, we multiply both sides by $(T_b - T_\infty)$ and set $q_p^*$ equal to the general heat transfer equation. This analysis assumes that the outer diameter of the pan reaches the ambient temperature.

$$q_p^*(T_b - T_\infty) = -k_p(2\pi r_c t_p)\frac{dT}{dr} = -\frac{(T_b - T_\infty)}{R_p} = q\ [W] \quad (S7.7)$$

where $T_b$ is the temperature of the bottom of the container. By simple inspection, we can see that

$$R_p = -\frac{1}{q_p^*} \quad (S7.8)$$

By solving Eq. (S7.4) using the thermal conductivity of aluminum, a pan thickness of 1 mm, and a pan radius of 5 cm, we can estimate that

$$R_p = 68.898\frac{K}{W} \quad (S7.9)$$

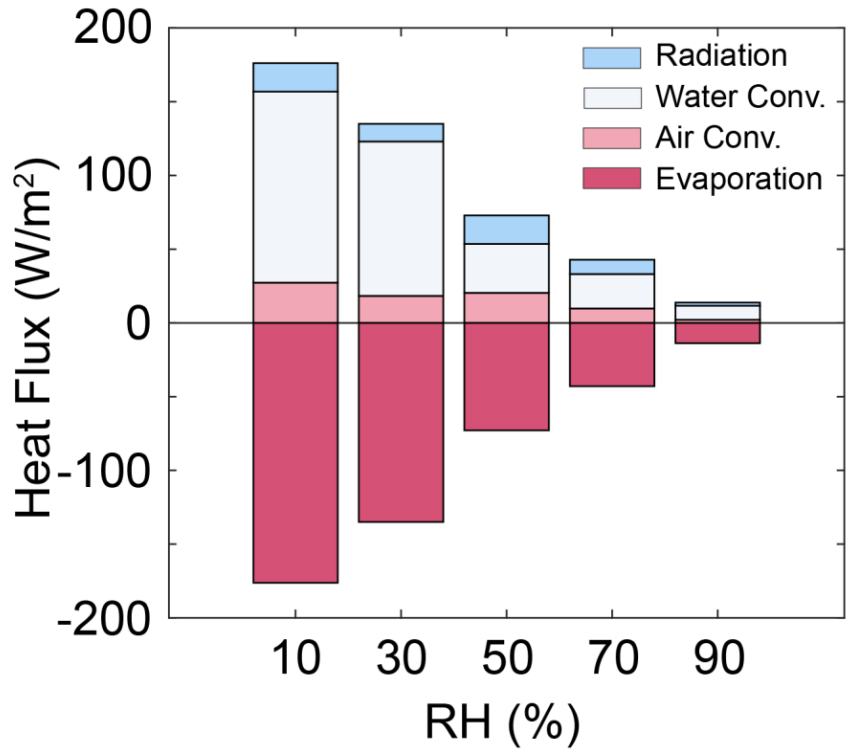

**Figure S1. Heat path for dark evaporation.** Heat fluxes from each of the pathways at the air-water interface in FEA simulations for a container with a diameter of 3 cm evaporating into an ambient at 23 °C.

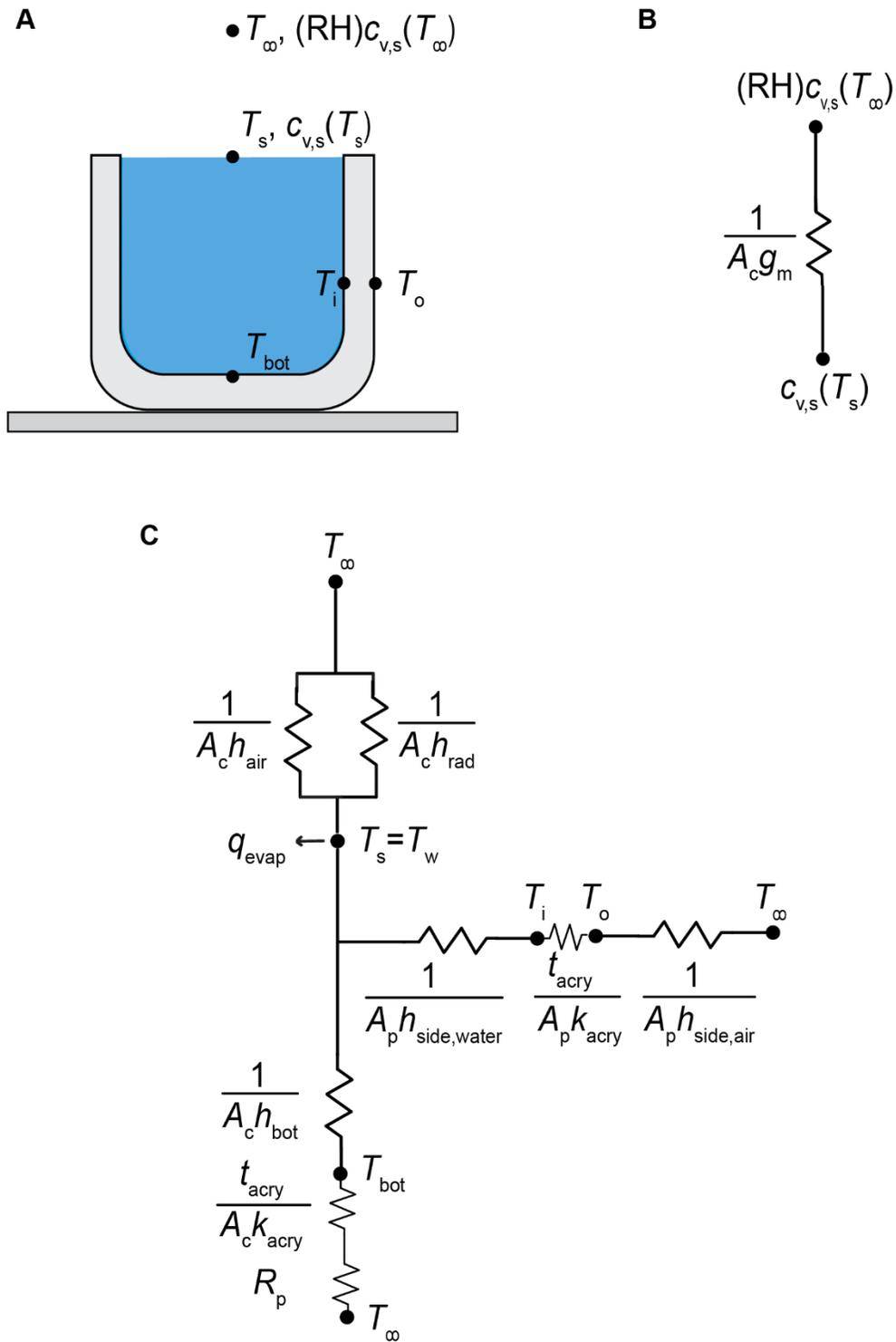

**Figure S2. Heat and mass transfer resistance network.** (**A**) Heat and mass transfer schematic for water-only natural evaporation with temperature and vapor concentration nodes. (**B**) Mass transfer resistance network and (**C**) heat transfer resistance network used to calculate water-only evaporation rates from SI Note 2.

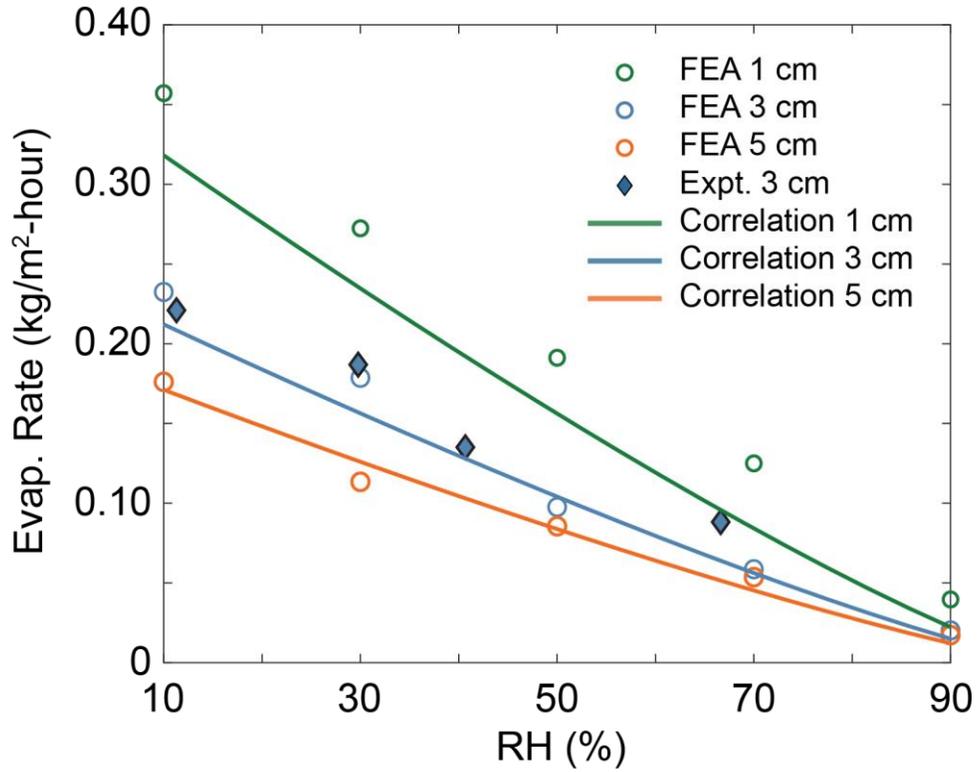

**Figure S3. Validation of correlation-based model.** Predicted evaporation rate for different sized container diameters using the correlation model described in SI Note 2 and Fig. S2. Comparisons with FEA simulations and experiments are shown as well.

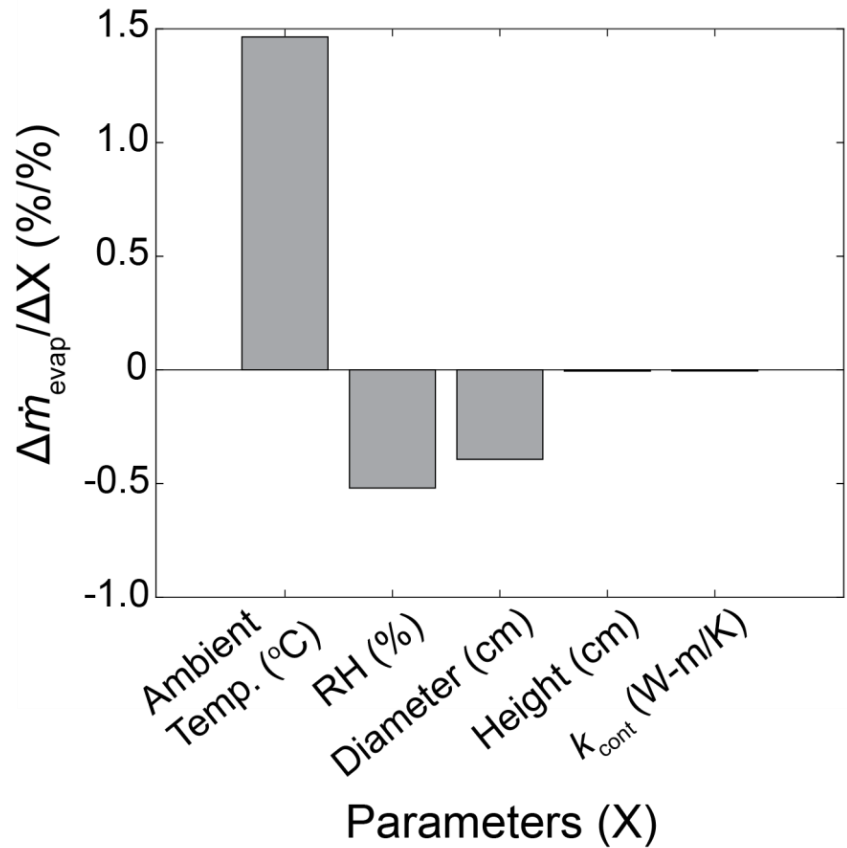

**Figure S4. Sensitivity analysis of the correlation model.** The sensitivity of evaporation is calculated numerically by changing the base case parameters by 0.5% in both the positive and negative directions and calculating the relative change in evaporation rate predicted. The base case considered is a water-only interface evaporating into an ambient at 23 °C, an ambient RH of 30%, an evaporating diameter of 3 cm, a container wall thickness of 2 mm, a container height of 4 cm, and a thermal conductivity of the container at 0.19 W/m-K.

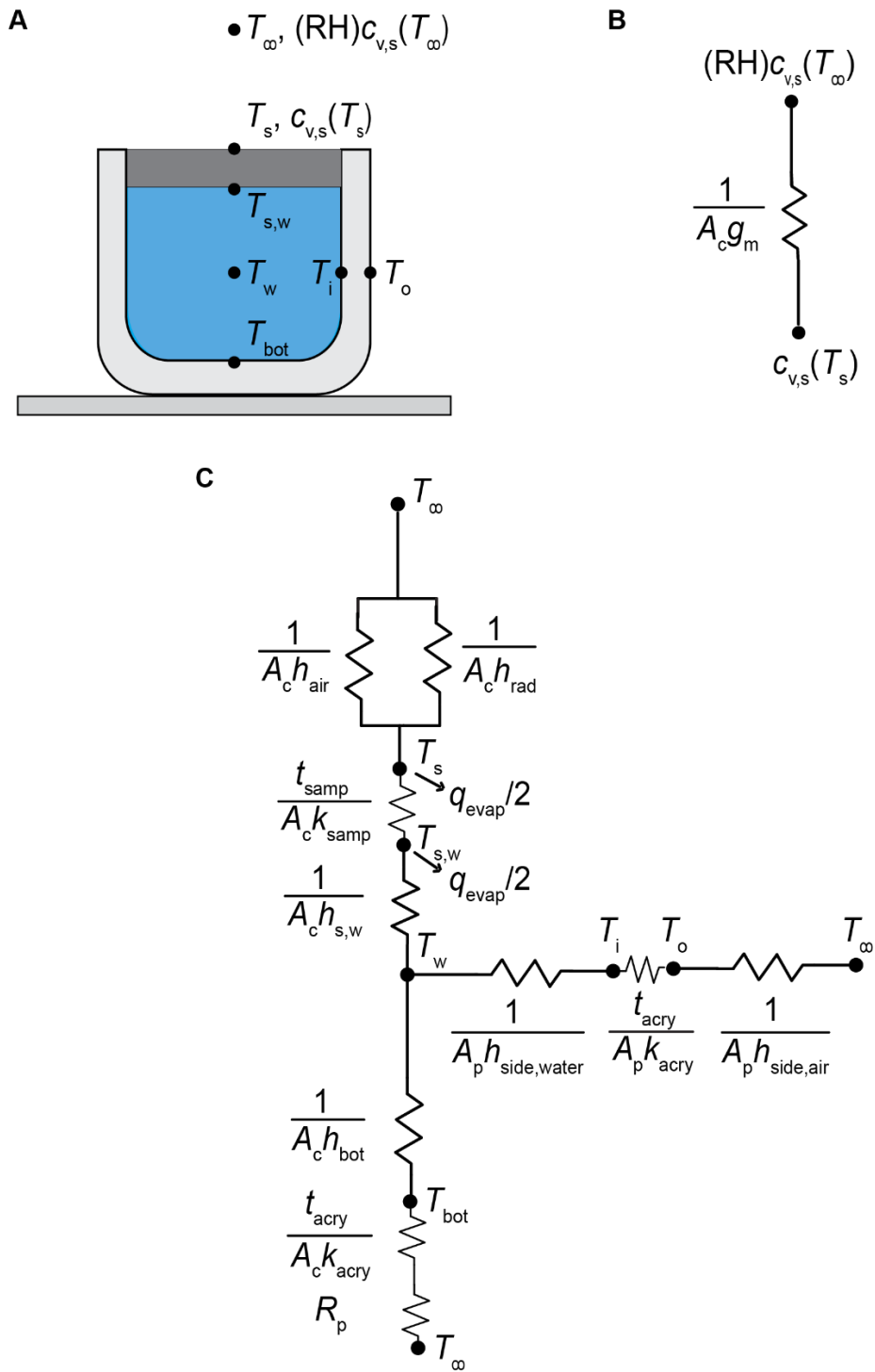

**Figure S5. Heat and mass transfer resistance network when there is latent heat reduction.** (**A**) Heat and mass transfer schematic for porous sample natural evaporation with temperature and vapor concentration nodes. (**B**) Mass transfer resistance network and (**C**) heat transfer resistance network used to calculate evaporation rates for porous evaporators described in SI Notes 3-6.

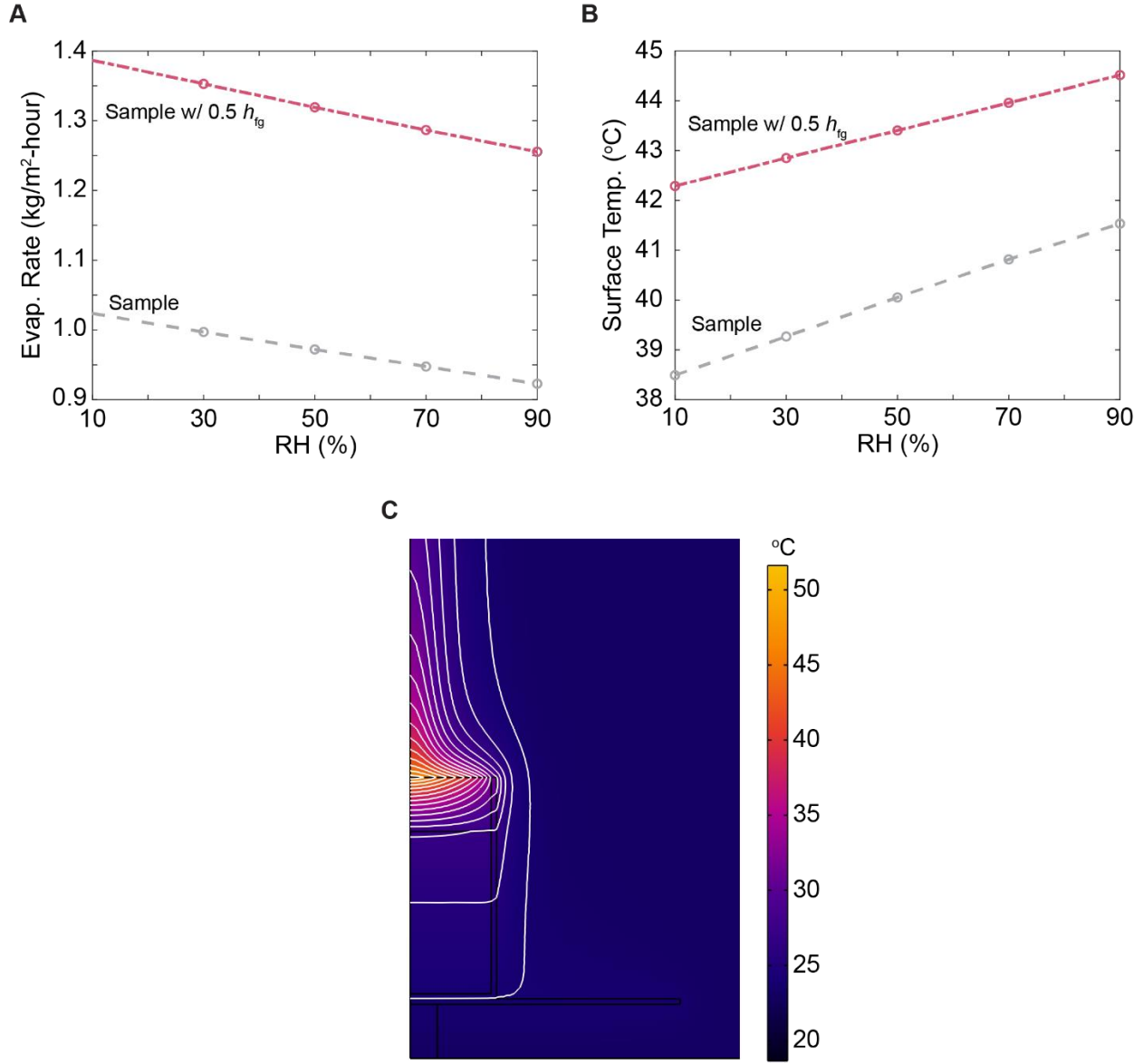

**Figure S6. FEA heat and mass transfer under solar irradiation.** Simulated (**A**) evaporation rates and (**B**) average surface temperatures of porous sample during solar evaporation experiments. The evaporating diameter is set to 3 cm, the ambient temperature is set to 23 °C, and the sample thermal resistance in the height direction is 0.1 m$^2$K/W. (**C**) Snapshot of steady-state temperature distribution after 2 hours of solar evaporation for a sample with no intermediate water states. The white lines represent isotherm contours.

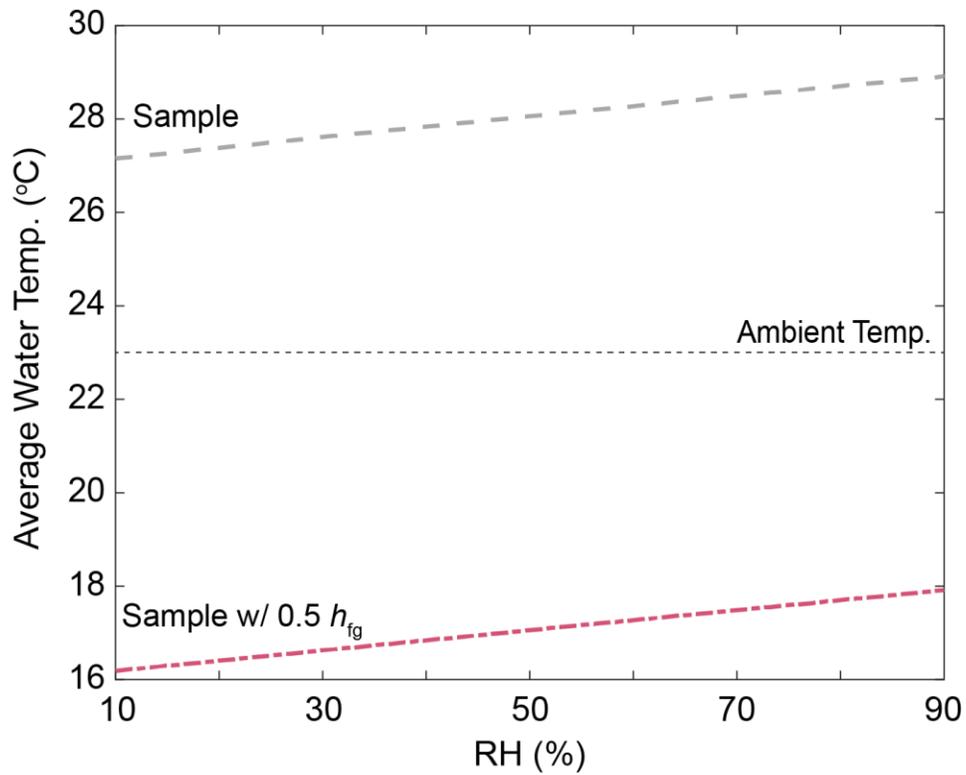

**Figure S7. Predicted average water temperature inside of the container under solar irradiation.** The container size is set to 3 cm, sample resistance of 0.1 m²K/W, and ambient temperature of 23 °C in the correlation model. If latent heat is reduced by half, the water temperature is lower than ambient due to cooling effect at the sample-water interface.

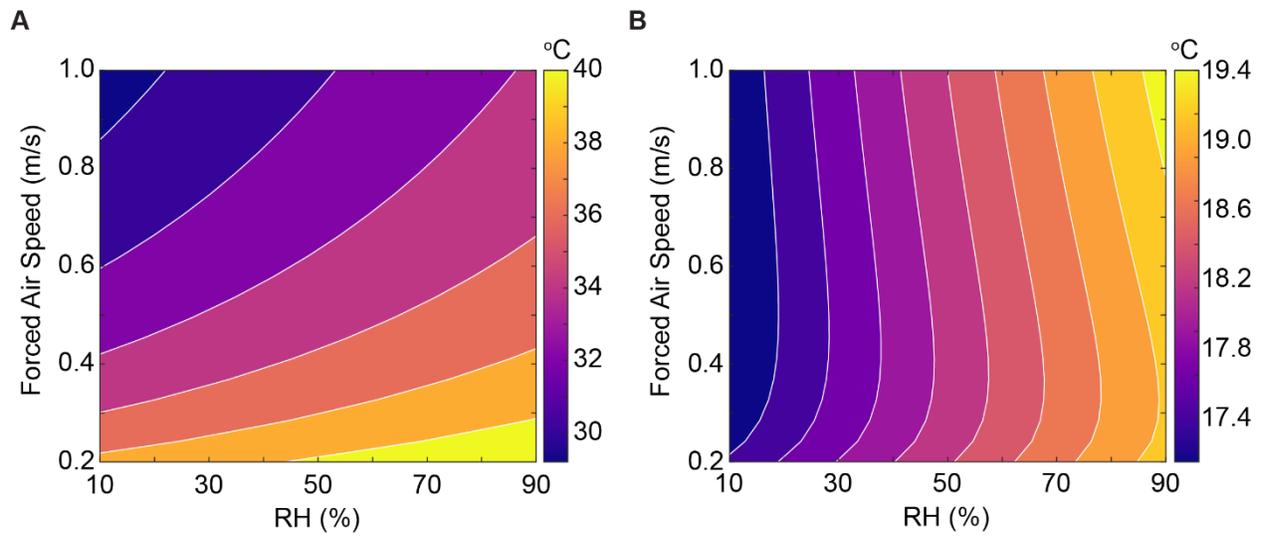

**Figure S8. Effect of forced convection.** Predicted (**A**) surface temperature and (**B**) average water temperature of from evaporation of porous evaporators with reduced latent heat (half of bulk value) during solar evaporation under forced convection conditions in correlation model. Water temperature is again below the ambient temperature at 23 °C.

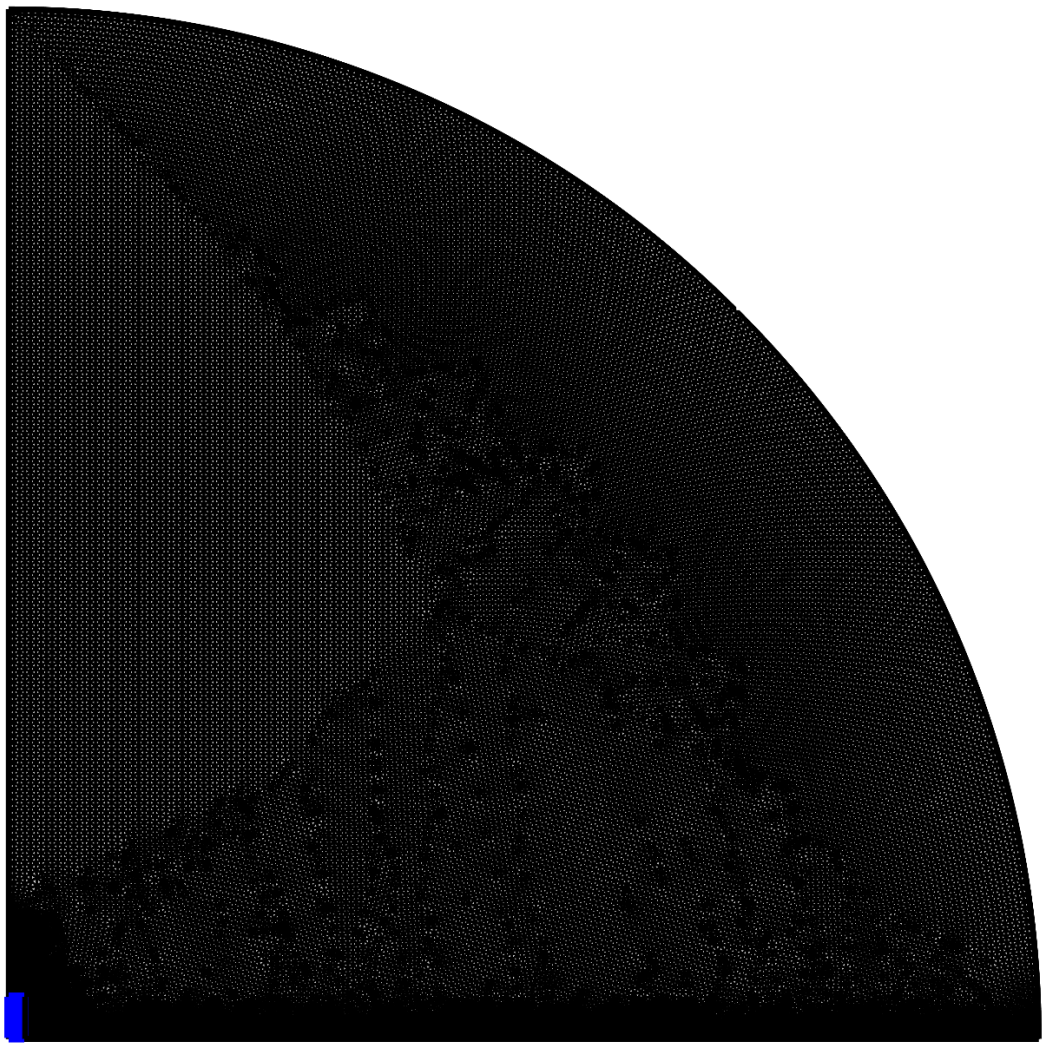

**Figure S9. Simulation domain.** Full simulation domain with 1 meter radius for natural evaporation of pure water interfaces. The mesh representing the water container and liquid water inside is highlighted in blue for clarity.

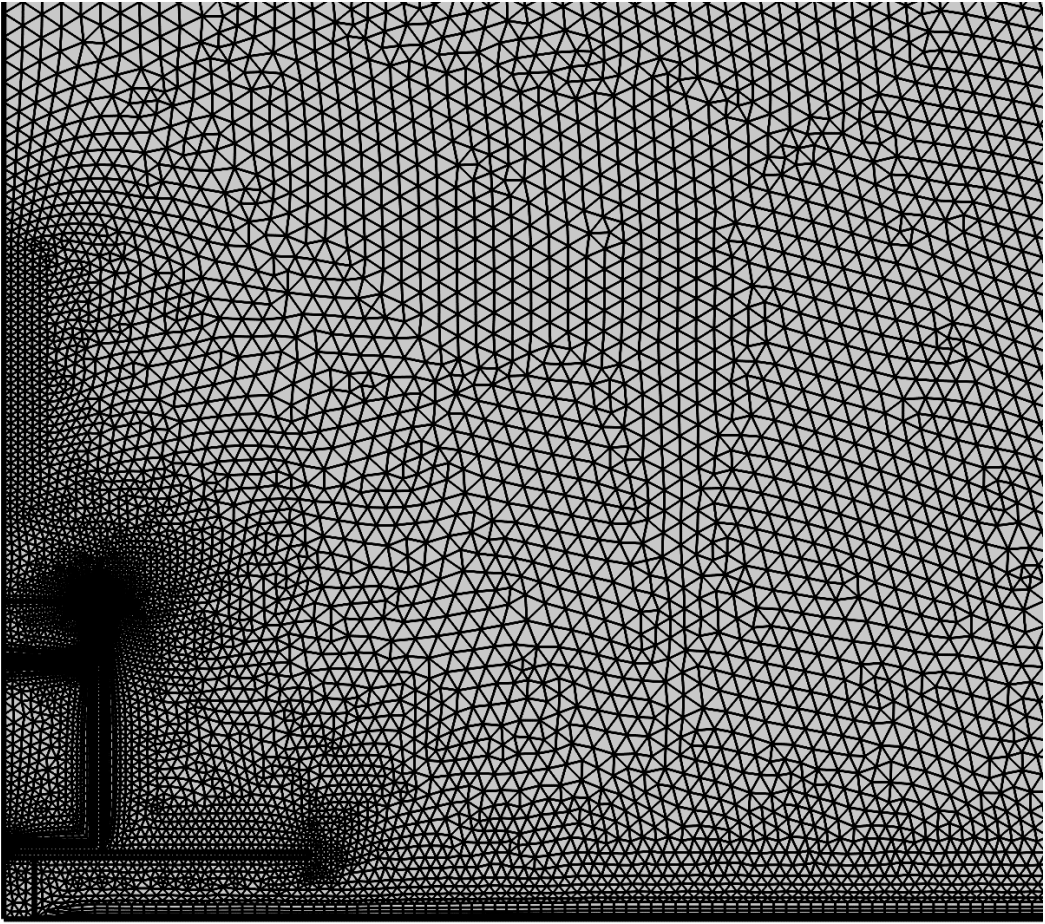

**Figure S10. Mesh used for solar interfacial-evaporation simulation.** Zoomed in mesh for solar evaporation of samples in FEA simulations on top of an aluminum scale pan. Higher resolution mesh is used in the boundary layer regions on the evaporating surface and near solid/fluid interfaces. The hemispherical air domain size is reduced to 0.6 m to reduce computation time.

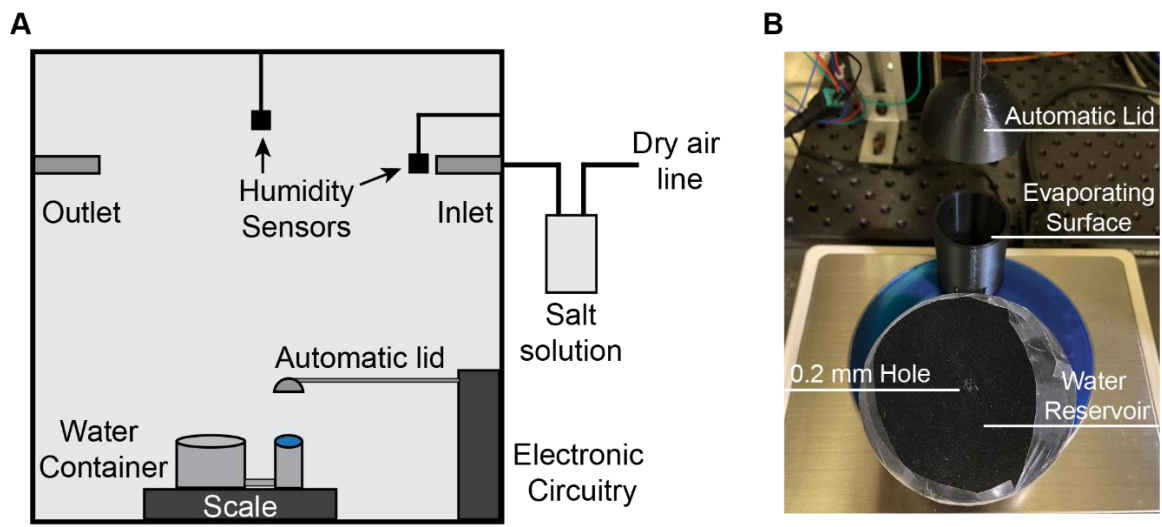

**Figure S11. Experimental set up.** (**A**) Schematic of controlled humidity chamber setup with a dimension of 0.6 m in each direction. (**B**) Top view of the 3D printed polylactic acid container with water on the scale and the automatic lid lifted above the surface.

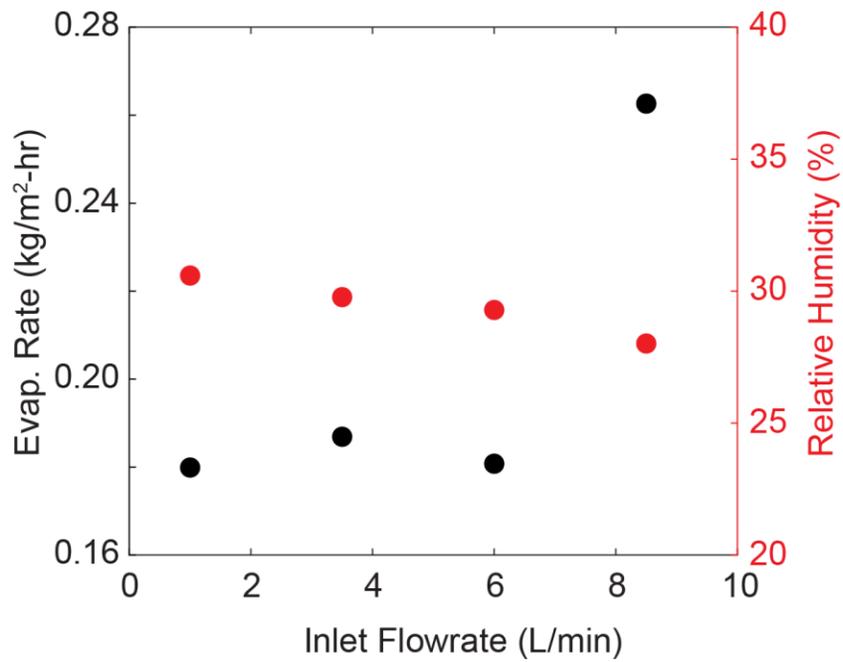

**Figure S12. Effect of inlet airflow.** Dependence on evaporation rate with inlet airflow for water-only natural evaporation experiments. The experiments were conducted with a mean temperature of 24.62 °C. The difference in mean temperatures for each experiment was below 0.12 °C.

**Table S1. Comparison of literature natural evaporation compared to model.** Data shows the experimental testing conditions as well as the sample geometry. These values are used as inputs into the correlation model to predict water-only evaporation rate as described in SI Note 3.

| Source | Size (cm) | Shape | RH (%) | Temp (°C) | Water Evap. Rate (kg/m²-hr) | Sample Evap. Rate (kg/m²-hr) | Predicted Water Evap. Rate (kg/m²-hr) |
|---|---|---|---|---|---|---|---|
| Study 1 (9) | 1 | Square | 45 | 25 | 0.011 | 0.019 | 0.201 |
| Study 2 (24) | 1 | Square | 45 | 25 | 0.014 | 0.039 | 0.201 |
| Study 3 (25) | 1.6 | Circle | 45 | 25 | 0.073 | 0.099 | 0.169 |
| Study 4 (30) | 3 | Circle | 43.5 | 31 | 0.416 | 0.502 | 0.192 |
| Study 5 (26) | 3.2 | Circle | 31 | 20 | 0.081 | 0.141 | 0.124 |
| Study 6 (27) | 3.5 | Square | 30 | 20 | 0.042 | 0.085 | 0.122 |
| Study 7 (14) | 4.1 | Circle | 18.6 | 19.55 | 0.084 | 0.113 | 0.133 |
| Study 8 (28) | 5 | Circle | 27 | 26 | 0.059 | 0.272 | 0.157 |
| Study 9 (29) | 5.5 | Square | 45 | 25 | 0.024 | 0.059 | 0.106 |

**Table S2. Comparison of this work's experimental data with correlation model.**

| Size (cm) | Shape | RH (%) | Temp (C) | Measured Water Evap. Rate (kg/m²-hr) | Predicted Water Evap. Rate (kg/m²-hr) | Rel. Error to Model |
|---|---|---|---|---|---|---|
| 3 | Circle | 11.30 | 24.83 | 0.221 | 0.233 | -5% |
| 3 | Circle | 29.77 | 24.57 | 0.187 | 0.173 | 8% |
| 3 | Circle | 40.67 | 24.66 | 0.135 | 0.142 | -5% |
| 3 | Circle | 66.63 | 24.74 | 0.088 | 0.071 | 24% |